\documentclass[reprint,floatfix]{revtex4-1}

\usepackage{amsmath}
\usepackage{booktabs} 
\usepackage{bm}
\usepackage{makecell}
\usepackage{graphicx}
\usepackage{xcolor}
\usepackage{adjustbox}
\begin{document}

\title{Ghost-Mode Filtered Fluctuating Lattice Boltzmann Method}

\author{M. Lauricella}
\email{marco.lauricella@cnr.it}
\thanks{Corresponding author}
\affiliation{Istituto per le Applicazioni del Calcolo CNR, via dei Taurini 19, 00185 Rome, Italy}
\author{A. Montessori}
\affiliation{Department of Civil, Computer Science and Aeronautical Technologies Engineering, Roma Tre University, Via Vito Volterra, Rome, 00146, Italy}

\author{A. Tiribocchi}
\affiliation{Istituto per le Applicazioni del Calcolo CNR, via dei Taurini 19, 00185 Rome, Italy}
\affiliation{INFN "Tor Vergata" Via della Ricerca Scientifica 1, 00133 Rome, Italy}
\author{S. Succi}
\affiliation{Center for Life Nano Science@La Sapienza, Istituto Italiano di Tecnologia, 00161 Roma, Italy}
\affiliation{Department of Physics, Harvard University, Cambridge, MA, 02138, USA}
\affiliation{Istituto per le Applicazioni del Calcolo CNR, via dei Taurini 19, 00185 Rome, Italy}

\date{\today}

\begin{abstract}
Fluctuating lattice Boltzmann solvers are widely employed to model mesoscopic fluid behavior in soft-matter systems, including colloidal suspensions and dilute polymer solutions. Despite their utility, these methods can lose accuracy and stability when non-hydrodynamic modes interfere with the dynamics, especially in single--relaxation-time schemes. Here, we introduce a ghost-mode filtered fluctuating lattice Boltzmann method (GMF-FLBM) for the D3Q27 lattice, obtained by selectively eliminating the propagation of the ghost deterministic content while preserving the necessary stochastic forcing. We show, over a broad range of relaxation times, that GMF-FLBM recovers the amplitudes of equilibrium fluctuations with a comparable accuracy as a fully regularized high-order formulation, while requiring only minor adjustments to the conventional BGK collision framework.
\end{abstract}

% insert suggested PACS numbers in braces on next line

\maketitle %\maketitle must follow title, authors, abstract and \pacs

\begin{center}
    \textit{
        This work is dedicated to Kurt Kremer, a highly esteemed colleague and pioneer of computer simulation, with the warmest wishes of great continued success for many years to come. 
    }
\end{center}

\section{Introduction}
\label{sec:introduction}

In the last decades, the lattice Boltzmann method (LBM) has gained a prominent role as a robust and versatile framework for tackling the Navier–Stokes equations within a kinetic formulation \cite{tiribocchi2025lattice,succi2018lattice,kruger2017lattice,sukop2006lattice}.
In particular, LBM was commonly employed across a broad range of applications such as flows in porous media \cite{boek2010lattice,guo2002lattice,cali1992diffusion}, phase transitions or phase separation driven by non-ideal interactions \cite{montessori2025thread,zhang2024improved,tiribocchi2020novel,chiappini2019hydrodynamic,montessori2019mesoscale,liu2012three,shan1993lattice}, electrohydrodynamic phenomena \cite{xiong2025thermodynamically,liu2024consistent,wang2021lattice,lauricella2018entropic,kupershtokh2006lattice}, particle-laden flows \cite{guglietta2023suspensions,yang2022capillary,bonaccorso2020lbsoft,harting2014recent,ladd2001lattice}, polymer dynamics \cite{monteferrante2021lattice,malaspinas2010lattice,berk2005lattice,ahlrichs1999simulation,ahlrichs1998lattice}, chemically reactive fluids \cite{sawant2021lattice,lin2017multi}, and various forms of active matter \cite{tiribocchi2023crucial,carenza2020chaotic,doostmohammadi2016stabilization,marenduzzo2007steady}, among many others. However, despite its broad success in modeling deterministic fluid dynamics, the traditional LBM framework is based on the assumption that thermal fluctuations can be safely ignored \cite{kruger2017lattice}. 

Neglecting thermal fluctuations is generally valid at macroscopic scales, but it breaks down as one approaches mesoscopic or nanometric regimes, where stochastic fluctuations become an essential ingredient of the fluid response \cite{adhikari2005fluctuating}. Under these conditions, incorporating noise in a thermodynamically 
consistent manner is no longer optional but crucial for capturing the correct physics.

A paradigmatic example is offered by dilute polymer solutions, where hydrodynamic fluctuations strongly influence chain dynamics, relaxation spectra, and even the diffusion-controlled reactive encounters of polymerization in solution. Simulations by Dünweg and Kremer demonstrated that a polymer in a fluctuating solvent exhibits Zimm–type dynamics emerging directly from the interplay between thermal noise and long–range hydrodynamic interactions \cite{dunweg1993molecular}. Subsequent multiscale developments, including hybrid MD–LB approaches and adaptive–resolution schemes, further highlighted the mechanism by which fluctuations mediate conformational sampling and transport in reactive polymeric systems \cite{ahlrichs1999simulation,praprotnik2008multiscale}. 

Over the last two decades, the fluctuating lattice Boltzmann method (FLBM)  has become one of the most versatile tools for simulating such phenomena at nanometric resolutions, where hydrodynamic fields interact non-trivially with non-hydrodynamic kinetic degrees of freedom \cite{xue2021lattice,parsa2020large,belardinelli2015fluctuating,gross2011modelling}. Early fluctuating LB formulations, most notably those introduced by Ladd and co-workers for colloidal suspensions \cite{ladd2001lattice,ladd1994numerical,ladd1993short}, injected noise only into the hydrodynamic subspace, yielding correct long-wavelength behavior but systematically underestimating fluctuations at intermediate and high wave numbers. This issue was later clarified by Adhikari et al. \cite{adhikari2005fluctuating} and Dünweg et al. \cite{dunweg2009lattice,dunweg2007statistical}, who showed that a fully consistent 
Fluctuation Dissipation Theorem (FDT) requires stochastic forcing across all non-conserved modes, including ghost modes, when represented in an MRT/Hermite-orthogonal basis. 

Although this procedure is formally consistent, the presence of ghost moments in the populations may generate residual artifacts at moderate and large wave numbers, especially when lattice anisotropies interact with the stochastic forcing. These effects become more evident when the FLBM is used alongside the single-relaxation BGK fluctuating Lattice Boltzmann (BGK-FLBM) approach. In standard BGK-FLBM schemes, these spurious components accumulate and propagate across time, affecting both stability and the spectral distribution of fluctuations, even in full-Hermite lattices such as D3Q27 \cite{lauricella2025regularized}. 

Regularization strategies originally introduced to suppress lattice artifacts in athermal LB models \cite{jacob2018new,coreixas2017recursive,mattila2017high,latt2006lattice} offer a natural pathway to constrain these modes. However, their integration with fluctuating LB formulations has been explored only recently \cite{lauricella2025regularized}. In the framework of the regularized fluctuating lattice Boltzmann model (Reg-FLBM), Lauricella et al. \cite{lauricella2025regularized} showed that using the full Hermite expansion in the D3Q27 scheme to reconstruct equilibrium and non-equilibrium populations, combined with a mode-by-mode relaxation in Hermite space, significantly improves the enforcement of the fluctuation–dissipation theorem, especially at low Mach numbers ($\mathrm{Ma} \le 0.1$) and in weakly compressible flows.

In this framework, the key observation motivating the present paper is that truncating the equilibrium distribution at second order in the Hermite expansion, while keeping a unit relaxation rate for all ghost modes, acts as an effective ghost-mode filter.
Indeed, in this two relaxation setup framework, ghost modes should not retain any deterministic memory of the previous timestep. Their role is purely statistical and should be restricted to carrying the appropriate thermal noise. This observation leads to a simple modification of the fluctuating regularized scheme in Reg-FLBM: after performing the Hermite projection, ghost modes can be entirely suppressed and replaced only by stochastic amplitudes consistent with the fluctuation--dissipation theorem. 

The result is a ghost-mode filtered FLBM (GMF-FLBM), in which the hydrodynamic dynamics are preserved, while the non-hydrodynamic sector is reduced to its minimal statistical content. This yields cleaner hydrodynamic fluctuation spectra and reduces mode coupling across the Hermite hierarchy.

The present GMF-FLBM approach retains all advantages of the regularized lattice Boltzmann framework: orthogonal Hermite moments, a clean separation between low-order (hydrodynamic) and higher-order (non-hydrodynamic) terms, and a more robust collision operator than BGK operator. At the same time, GMF-FLBM avoids artifacts arising from the propagation of ghost-mode populations, providing the correct statistics of fluctuating hydrodynamics, even when the equilibrium distribution is truncated at second order in the Hermite expansion.

The paper is organized as follows. Section II outlines the construction of the ghost-mode filtered FLBM on the D3Q27 lattice. Section III presents equilibrium and non-equilibrium tests assessing the accuracy of the method. Section IV summarizes the main findings and outlines perspectives of possible applications.

\section{Method}
\label{sec:method}

In discrete velocity phase space, the  distribution function evolves according to the fluctuating lattice Boltzmann equation:
\begin{equation}
\begin{split}
f_i(x_\alpha + c_{i\alpha}\Delta t, t+\Delta t)
&= f_i^{\mathrm{eq}}(x_\alpha,t)+ (1-\omega)\,f_i^{\mathrm{neq}}(x_\alpha,t)  \\
&\quad +w_i \sum_{k=4}^{26} b_k^{-1} e_{ki} \varphi_k r_k,
\label{eq:flbe}
\end{split}
\end{equation}
where $f_i^{\mathrm{neq}} = (f_i - f_i^{\mathrm{eq}})$, and $f_i$ denotes the single--particle distribution (or population) associated with direction $i$ at position $x_\alpha$ and time $t$, according to the D3Q27 scheme, while $w_i$ are the lattice weights reported in Table \ref{tab:d3q27}, \( r_k \) is a standard normal random variable and $\varphi_k$  denotes the noise amplitude. In particular, the noise amplitude is assessed as \cite{dunweg2007statistical,schiller2008thermal,dunweg2009lattice}:
\begin{equation} \label{fluct_phi}
\varphi_k = \sqrt{ \frac{\rho\, k_B T\, \omega (2 - \omega)\, b_k}{c_s^2} },
\end{equation}
and it is chosen to ensure the thermodynamic consistency via the fluctuation--dissipation theorem with \( k_B T \) tuning the fluctuation variance of the non-conserved modes. The symbol $b_k$ denotes the normalization factor, while $e_{ki}$ is the vector of the orthogonal full--Hermite basis set specifically constructed for the D3Q27 lattice~\cite{malaspinas2015increasing} reported in Table \ref{TAB:HERMITE_D3Q27_COMPACT}.

The parameter $\omega = 1/\tau$ is the relaxation frequency, here taken for simplicity as a single scalar relaxation time as in the standard BGK model. 
It sets the kinematic viscosity through $\nu = c_s^2(\tau - 0.5)$, with $c_s^2 = 1/3$ in lattice units. 
The equilibrium distribution $f_i^{\text{eq}}$ is obtained as a discrete-velocity expansion of the Maxwell--Boltzmann distribution~\cite{succi2018lattice,kruger2017lattice}.

The conserved moments (hydrodynamic quantities) are computed as zeroth and first moments of the populations
\begin{align}
\rho &= \sum_i f_i(x_\alpha, t), \label{eq:rho} \\
\rho u_\alpha &= \sum_i f_i(x_\alpha, t) c_{i\alpha}, \label{eq:u}
\end{align}
where ${\bf u}$ is the fluid velocity. 

\begin{table}[h]
\centering
\caption{Discrete velocities and weights for the D3Q27 lattice.}
\label{tab:d3q27}
\begin{tabular}{c c c c}
\hline
\textbf{i} & \textbf{Velocity} $(c_{ix}, c_{iy}, c_{iz})$ & $\|\mathbf{c}_i\|^2$ & $w_i$ \\
\hline
0 & $(0,0,0)$ & 0 & $8/27$ \\
1--6 & $(\pm1,0,0), (0,\pm1,0), (0,0,\pm1)$ & 1 & $2/27$ \\
7--18 & $(\pm1,\pm1,0), (\pm1,0,\pm1), (0,\pm1,\pm1)$ & 2 & $1/54$ \\
19--26 & $(\pm1,\pm1,\pm1)$ & 3 & $1/216$ \\
\hline
\end{tabular}
\end{table}

\begin{table*}[ht]
\caption{Compact representation of the discrete Hermite basis for the D3Q27 lattice \cite{lauricella2025regularized,malaspinas2015increasing}.
The symbol $e_{ki}$ denotes the corresponding basis functions of the Hermite polynomial, $H^{(n)}_{\dots}$, of order $n_k$
while $\mu_k$ and $b_k$ are the multiplicity and normalization factor, respectively.
Hermite polynomials sharing the same values of  $n_k$, $\mu_k$ and $b_k$ are grouped together with $k$ running from 0 to 26 indicating the $k$-th basis vector.
The components of the basis functions $e_{ki}$ are denoted by $\alpha$, $\beta$, $\gamma$, etc., corresponding to the Cartesian indices written as subscripts of each symbol $\mathcal{H}_{\alpha_1 \ldots \alpha_n}^{(n_k)} $. Cartesian subscripts $x$, $y$, and $z$ denote specific components (no Einstein summation is implied).}
\centering
\small
\begin{adjustbox}{max width=\textwidth}
\begin{tabular}{cccccc}
\hline
$k$ & Hermite symbols $\mathcal{H}_{\alpha_1 \ldots \alpha_n}^{(n_k)}$ & $e_{ki}$ & $n_k$ & $\mu_k$ & $b_k$ \\
\hline
0 & $H^{(0)}$ & $1$ & 0 & 1 & $1$ \\
1--3 &
$H^{(1)}_{x}$, $H^{(1)}_{y}$, $H^{(1)}_{z}$ &
$c_{i\alpha}$ & 1 & 1 & $1/3$ \\
4--6 &
$H^{(2)}_{xx}$, $H^{(2)}_{yy}$, $H^{(2)}_{zz}$ &
$c_{i\alpha}c_{i\beta} - c_s^2 \delta_{\alpha\beta}$ & 2 & 1 & $2/9$ \\
7--9 &
$H^{(2)}_{xy}$, $H^{(2)}_{xz}$, $H^{(2)}_{yz}$ &
$c_{i\alpha}c_{i\beta} - c_s^2 \delta_{\alpha\beta}$ & 2 & 2 & $1/9$ \\
10--15 &
$H^{(3)}_{x^2y}$, $H^{(3)}_{x^2z}$, $H^{(3)}_{y^2x}$,
$H^{(3)}_{z^2x}$, $H^{(3)}_{y^2z}$, $H^{(3)}_{z^2y}$ &
$c_{i\alpha}^2 c_{i\beta} - c_s^2 c_{i\beta}$ & 3 & 3 & $2/27$ \\
16 &
$H^{(3)}_{xyz}$ &
$c_{i\alpha} c_{i\beta} c_{i\gamma}$ & 3 & 6 & $1/27$ \\
17--19 &
$H^{(4)}_{x^2y^2}$, $H^{(4)}_{x^2z^2}$, $H^{(4)}_{y^2z^2}$ &
$c_{i\alpha}^2 c_{i\beta}^2 - c_s^2 (c_{i\alpha}^2 + c_{i\beta}^2) + c_s^4$
& 4 & 6 & $4/81$ \\
20--22 &
$H^{(4)}_{xyz^2}$, $H^{(4)}_{xzy^2}$, $H^{(4)}_{yzx^2}$ &
$c_{i\alpha} c_{i\beta} c_{i\gamma}^2 - c_s^2 c_{i\alpha} c_{i\beta}$
& 4 & 12 & $2/81$ \\
23--25 &
$H^{(5)}_{x^2yz^2}$, $H^{(5)}_{x^2zy^2}$, $H^{(5)}_{y^2xz^2}$ &
$c_{i\alpha}^2 c_{i\beta} c_{i\gamma}^2 - c_s^2 (c_{i\alpha}^2 c_{i\beta} + c_{i\beta} c_{i\gamma}^2) + c_s^4 c_{i\beta}$
& 5 & 30 & $4/243$ \\
26 &
$H^{(6)}_{x^2y^2z^2}$ &
$c_{i\alpha}^2 c_{i\beta}^2 c_{i\gamma}^2
 - c_s^2 (c_{i\alpha}^2 c_{i\beta}^2 + c_{i\alpha}^2 c_{i\gamma}^2 + c_{i\beta}^2 c_{i\gamma}^2)
 + c_s^4 (c_{i\alpha}^2 + c_{i\beta}^2 + c_{i\gamma}^2)
 - c_s^6$
& 6 & 90 & $8/729$ \\
\hline
\end{tabular}
\end{adjustbox}
\label{TAB:HERMITE_D3Q27_COMPACT}
\end{table*}

The full-Hermite
D3Q27 representation allows both parts of the distribution to be written on the discrete Hermite basis:
\begin{equation}
f_i^{\mathrm{eq}} 
= 
w_i \sum_{k}
\frac{\mu_k}{c_s^{2n_k} \, n_k!}\,
\mathcal{H}_{i, \alpha_1 \ldots \alpha_n}^{(n_k)}  a^{(n_k)}_{\mathrm{eq}, \alpha_1 \ldots \alpha_{n_k}},
\label{eq:expeq}
\end{equation}
\begin{equation}
f_i^{\mathrm{neq}} 
= 
w_i \sum_{k}
\frac{\mu_k}{c_s^{2n_k} \, n_k!}\,
\mathcal{H}_{i, \alpha_1 \ldots \alpha_{n_k}}^{(n_k)}  a^{(n_k)}_{\mathrm{neq}, \alpha_1 \ldots \alpha_{n_k}},
\label{eq:expneq}
\end{equation}
where $\mathcal{H}_i^{(n_k)}$ denotes the $n_k$-th order
discrete Hermite tensor reported in Table \ref{TAB:HERMITE_D3Q27_COMPACT} with the indices \( \alpha_1 \ldots \alpha_n \) of the Hermite polynomials running over the spatial directions \( x, y, z \), while $\mu_k$ denotes the multiplicity of a Hermite polynomial given by the number of distinct tensors generated by permuting its indices. 
Hence, the normalization factors $b_k$ are connected to the orthogonal full--Hermite basis through the relation $b_k = (n_k! \cdot c_s^{2n_k})/\mu_k$ as reported in Ref. \cite{lauricella2025regularized}.

In Eq.s \eqref{eq:expeq} and \eqref{eq:expneq}, the symbols $a^{(n_k)}_{\mathrm{eq}, \alpha_1 \ldots \alpha_{n_k}}$ and $a^{(n_k)}_{\mathrm{neq}, \alpha_1 \ldots \alpha_{n_k}}$ denote the associated Hermite coefficients:
\begin{equation}
a^{(n_k)}_{\mathrm{eq}, \alpha_1 \ldots \alpha_{n_k}}
=
\sum_i f_i^{\mathrm{eq}}\,
H_{i, \alpha_1 \ldots \alpha_{n_k}}^{(n_k)},
\end{equation}
\begin{equation}
a^{(n_k)}_{\mathrm{neq},\alpha_1 \ldots \alpha_{n_k}}
=
\sum_i f_i^{\mathrm{neq}}\,
H_{i, \alpha_1 \ldots \alpha_{n_k}}^{(n_k)},
\label{eq:coeffneq}
\end{equation}
so that the full set of Hermite coefficients is obtained by projecting $f_i^{\mathrm{eq}}$ and $f_i^{\mathrm{neq}}$ up to the highest order supported by the lattice.

A key point of the regularization procedure is that the non–equilibrium part $f_i^{\mathrm{neq}}$ is not
used directly as $(f_i - f_i^{\mathrm{eq}})$. Instead, it is reconstructed by projecting the
distribution onto the Hermite basis reported in Table \ref{TAB:HERMITE_D3Q27_COMPACT}.
Further, the reconstruction step provides a clean separation between hydrodynamic modes and
higher–order ghost modes, allowing a direct control on the highest Hermite order retained in the reconstruction of both equilibrium and non-equilibrium population terms.

In this framework, the filtering step is introduced by restricting the Hermite expansion to the
subset of modes one wishes to retain. In practice, keeping only the Hermite tensors up to second
order, corresponding to the hydrodynamic sector, we obtain a ghost–filtered fluctuating lattice Boltzmann
scheme, where all higher–order ghost contributions are removed by construction and only the
hydrodynamic tensors are kept. In the
collision operator this corresponds to relaxing the non–equilibrium ghost modes with
$\omega = 1$, since the post–collision term multiplies the non--equilibrium contribution by $(1-\omega)$. The noise
acting on the ghost sector ($k$ indices $10$--$26$), denoted $\varphi_k^g$, therefore involves a unit relaxation
\begin{equation} \label{fluct_phig}
\varphi_k^g = \sqrt{ \frac{\rho\, k_B T\, \, b_k}{c_s^2} },
\end{equation}
whereas the non–equilibrium hydrodynamic modes ($k$ indices $4$--$9$) continue to relax with the physical rate $\omega$ according to Eq. \eqref{fluct_phi}.

Finally, the ghost-mode filtered Fluctuating Lattice Boltzmann equation reads:
\begin{equation}
\begin{split}
f_i(x_\alpha + c_{i\alpha}\Delta t, t+\Delta t)
&= f_i^{\mathrm{eq}}(x_\alpha,t)
 + (1-\omega)\, f_i^{\mathrm{neq}}(x_\alpha,t) \\[4pt]
&\quad + w_i \sum_{k=4}^{9}  b_k^{-1} e_{ki}\,\varphi_k r_k \\[2pt]
&\quad + w_i \sum_{k=10}^{26} b_k^{-1} e_{ki}\,\varphi_k^{g} r_k ,
\end{split}
\label{eq:flbe2}
\end{equation}
where both the equilibrium, $f_i^{\mathrm{eq}}$, and non–equilibrium, $f_i^{\mathrm{neq}}$, terms are assessed at second order in the Hermite expansion by Eq.s \eqref{eq:expeq} and \eqref{eq:expneq} as:
\begin{equation}
f_i^{\mathrm{eq}} = w_i \rho \left( 1 + \frac{c_{i\alpha} u_\alpha}{c_s^2} + \frac{(c_{i\alpha}c_{i\beta} - c_s^2 \delta_{\alpha\beta}) u_\alpha u_\beta}{2 c_s^4} \right)
\end{equation}
\begin{equation}
f_i^{\mathrm{neq}}=w_i\frac{(c_{i\alpha}c_{i\beta} - c_s^2 \delta_{\alpha\beta}) \,a^{(2)}_{\mathrm{neq},\alpha\beta}}{2 c_s^4}.
\end{equation}
Here, repeated Greek indices imply summation (Einstein convention), and the second order Hermite coefficient, $a^{(2)}_{\mathrm{neq},\alpha\beta}$, is assessed according to Eq. \eqref{eq:coeffneq}, namely:
\begin{equation} \label{moments2}
    a^{(2)}_{\mathrm{neq},\alpha\beta}=\sum_i (f_i - f_i^{\mathrm{eq}})\,(c_{i,\alpha} c_{i,\beta} - c_s^2 \delta_{\alpha\beta}).
\end{equation}

It is worth noting that this formulation is thread–safe, since the evaluation of the equilibrium and non–equilibrium terms relies only on the local hydrodynamic fields, whereas the update of the populations is performed as a separate write operation. 
In this way, each thread of a GPU device reads exclusively macroscopic quantities and writes to its own lattice site, avoiding race conditions associated with non–local access to the population vector~\cite{montessori2023thread,montessori2024high}.

\section{Results}
\label{sec:results}

\begin{figure*}[t]
  \centering
\includegraphics[width=0.95\textwidth]{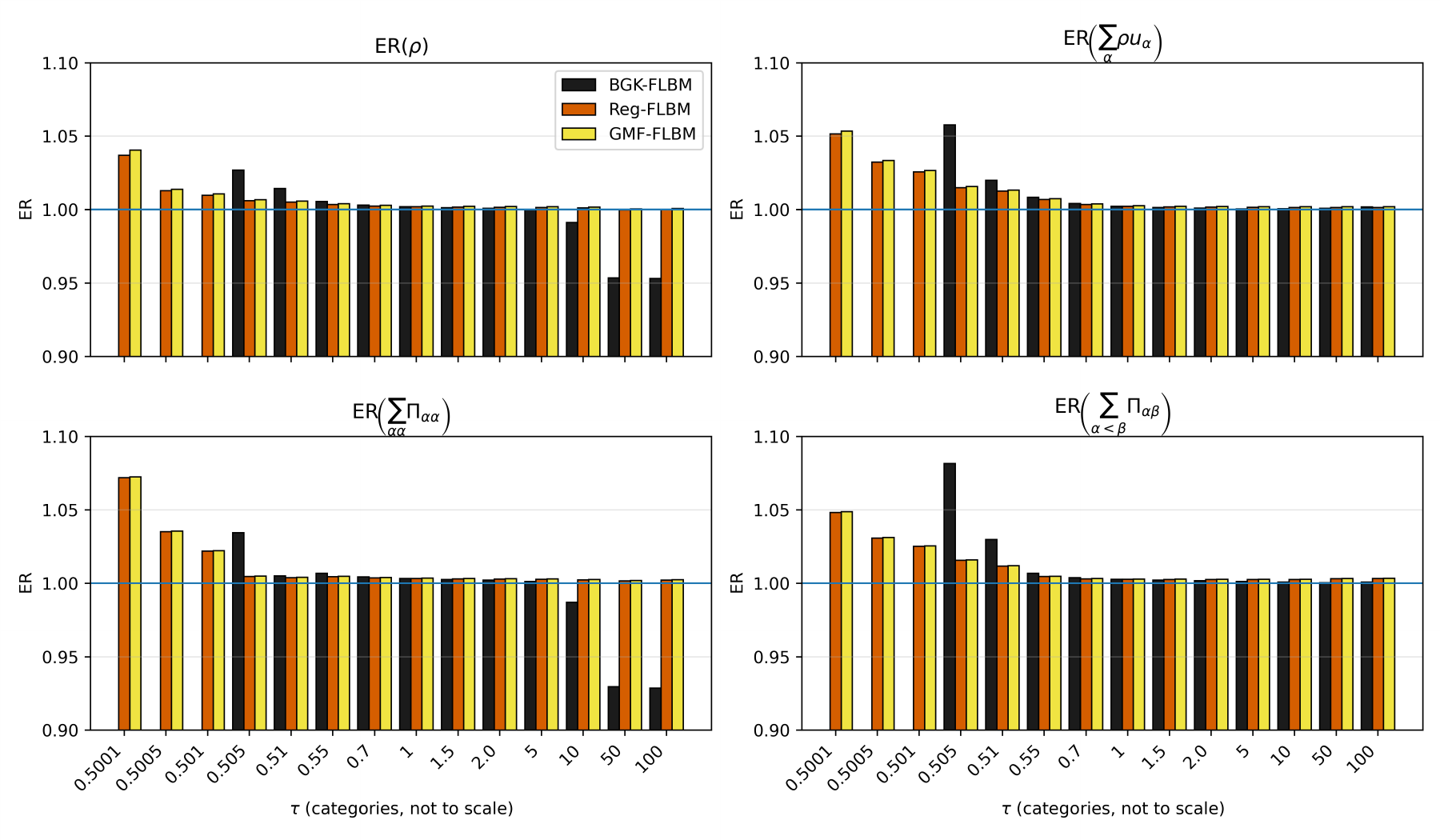}
  \caption{Equilibration ratio (ER) for four hydrodynamic observables as a function of the relaxation time $\tau$. The panels report $\mathrm{ER}(\rho)$, $\mathrm{ER}\!\left(\sum_\alpha \rho u_\alpha\right)$, $\mathrm{ER}\!\left(\sum_{\alpha\alpha}\Pi_{\alpha\alpha}\right)$, and $\mathrm{ER}\!\left(\sum_{\alpha<\beta}\Pi_{\alpha\beta}\right)$ (from top left to bottom right). For each selected $\tau$ (shown as discrete categories, not to scale) bars compare BGK-FLBM, Reg-FLBM, and GMF-FLBM; the horizontal line marks the target value $\mathrm{ER}=1$. Missing BGK-FLBM bars correspond to runs that could not be completed due to numerical instabilities of the BGK collision operator at those $\tau$ values.}
  \label{fig:er_comparison}
\end{figure*}

\begin{figure*}[t]
  \centering
\includegraphics[width=0.95\textwidth]{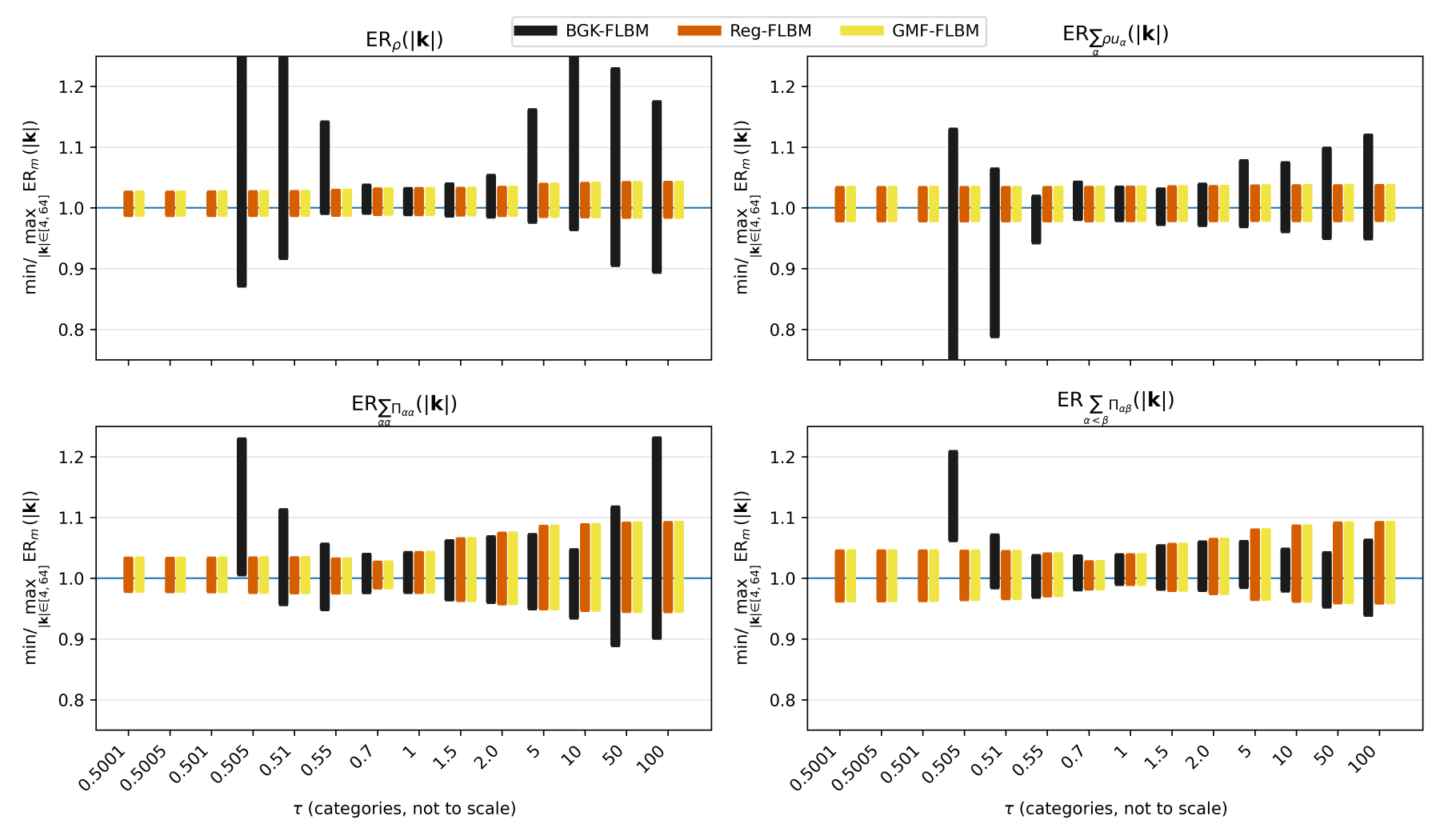}
\caption{Minimum and maximum spectral equilibration ratios for BGK-FLBM, Reg-FLBM, and GMF-FLBM.
For each $\tau$ and observable $m\in\{\rho,\ \sum_\alpha \rho u_\alpha,\ \sum_\alpha \Pi_{\alpha\alpha},\ \sum_{\alpha<\beta}\Pi_{\alpha\beta}\}$,
we compute the spherically averaged spectrum $S_m(|\mathbf{k}|)$ and form
$\mathrm{ER}_m(|\mathbf{k}|)=S_m^{\mathrm{num}}(|\mathbf{k}|)/S_m^{\mathrm{th}}(|\mathbf{k}|)$.
The plotted vertical segments (with horizontal end-caps) span the minimum and maximum of $\mathrm{ER}_m(|\mathbf{k}|)$ over $|\mathbf{k}|\in[4,64]$.
The horizontal line marks $\mathrm{ER}_m=1$; missing BGK-FLBM segments indicate unavailable $\tau$ cases.}
\label{fig:sminmax_comparison}
\end{figure*}

The ghost-mode filtered fluctuating lattice Boltzmann method (GMF-FLBM) was implemented in the multi-GPU \textit{accLB} code \cite{lauricella2025acclb}.
Hence, we ran a series of simulations on a cubic lattice with $256^3$ nodes to assess the implementation of the three different models. Mass, length, and time units were chosen so that the density satisfies $\rho = 1$ on a unit lattice, while the kinematic viscosity was varied by tuning the hydrodynamics relaxation time $\tau=1/\omega$ from 0.5001 (vanishing viscosity regime) to 100 (high viscosity fluids such as concentrated polymer solutions). This broad range of viscosities spans both the over-relaxed (\(1 < \omega < 2\)) and strongly under-relaxed (\(0 < \omega < 1\)) hydrodynamic regimes, thus providing a stringent test of the fluctuating models across distinct relaxation conditions.

Each run was advanced for 500{,}000 time steps, and $\rho$, the momentum field $\rho \boldsymbol{u}$, and the components of the momentum-flux tensor $\Pi_{\alpha\beta}$ were stored every 5{,}000 steps, giving 100 temporal snapshots per simulation. All simulations were performed at a fixed thermal energy \(k_B T = 1/3000\), consistent with earlier works~\cite{adhikari2005fluctuating,dunweg2007statistical}.

To better highlight the efficiency of the GMF-FLBM, we compare its results with those obtained using the standard BGK fluctuating lattice Boltzmann scheme (BGK-FLBM) and the high-order regularized fluctuating lattice Boltzmann model (Reg-FLBM) at the same operative conditions reported in our previous work~\cite{lauricella2025regularized}. In particular, the BGK-FLBM exploits Eq.~\eqref{eq:flbe} with a single relaxation frequency, while the Reg-FLBM is based on Eq.~\eqref{eq:flbe2}, together with the complete high-order expansion given in Eqs.~\eqref{eq:expeq} and~\eqref{eq:expneq} of both the equilibrium, \(f_i^{\mathrm{eq}}\), and non-equilibrium, \(f_i^{\mathrm{neq}}\), over the full Hermite basis set for the D3Q27 scheme reported in Table~\ref{TAB:HERMITE_D3Q27_COMPACT}.

We first quantify the quality of the thermalization of fluctuating hydrodynamic fields through the \emph{equilibration ratio} (ER), defined as the ratio between the variance measured in the simulation and the corresponding equilibrium prediction from statistical mechanics~\cite{landau1987fluid}. In this way, ER provides a direct metric of how faithfully the numerical scheme reproduces thermal fluctuations~\cite{adhikari2005fluctuating}. We define
\begin{equation}
\mathrm{ER}(m)=
\frac{\langle (\delta m)^2 \rangle}{\langle (\delta m)^2 \rangle_{\mathrm{theory}}},
\end{equation}
where $m$ is a fluctuating mode (e.g., density, momentum, or a stress component), $\delta m = m-\langle m\rangle$, $\langle (\delta m)^2\rangle$ is the variance computed from the simulation, and $\langle (\delta m)^2\rangle_{\mathrm{theory}}$ is the equilibrium value provided by the fluctuation--dissipation framework~\cite{landau1987fluid}. Therefore, $\mathrm{ER}=1$ signals exact agreement with equilibrium statistical mechanics, whereas departures from unity indicate imperfect equilibration.

Figure~\ref{fig:er_comparison} summarizes the equilibration ratios of four hydrodynamic observables, density ($\mathrm{ER}(\rho)$), total momentum ($\mathrm{ER}(\sum_\alpha \rho u_\alpha)$), diagonal momentum flux tensor ($\mathrm{ER}(\sum_{\alpha} \Pi_{\alpha\alpha})$), and off-diagonal momentum flux tensor ($\mathrm{ER}(\sum_{\alpha<\beta} \Pi_{\alpha\beta})$), over a broad range of relaxation times. For the regularized and ghost-mode filtered schemes, the ER values remain close to unity for all $\tau$ considered, with only a mild overestimation at the smallest viscosities. In particular, Reg-FLBM and GMF-FLBM yield nearly indistinguishable results, and both rapidly approach $\mathrm{ER}\simeq 1$ already for $\tau \gtrsim 0.55$.

In contrast, the BGK-FLBM shows a markedly less robust behavior. Close to the stability limit, it exhibits pronounced departures from equilibrium, most clearly in the momentum-related quantities and in the off-diagonal stress components, where the deviation peaks around $\tau=0.505$. At larger relaxation times the BGK results do not collapse to the same accuracy as the other two models and display a systematic underestimation of the density and diagonal-stress fluctuations at the highest $\tau$ values reported. For several $\tau$ values the BGK runs could not be completed because the collision step becomes numerically unstable, and the corresponding bars are therefore absent. Overall, the figure highlights that regularization and ghost-mode filtering substantially improve the fidelity and robustness of the fluctuating dynamics, yielding consistently thermalized fluctuations across the explored range of $\tau$.

An additional point emerging from Fig.~\ref{fig:er_comparison} is the close agreement between GMF-FLBM and Reg-FLBM across the entire range of relaxation times. Even at the most demanding values of $\tau$, where deviations from $\mathrm{ER}=1$ are largest, the ghost-mode filtered scheme does not exhibit a noticeably larger error than the fully regularized high--order formulation. This is a relevant practical outcome: GMF-FLBM attains essentially the same level of accuracy in the equilibrium fluctuations while avoiding the explicit evaluation of the higher-order terms in the Hermite expansion. In other words, the reduction in computational complexity provided by ghost-mode filtering does not translate into a meaningful loss of precision in the reproduced thermal statistics.

To verify that thermal noise is reproduced not only in its overall strength but also mode by mode, we analyze fluctuations in Fourier space, where each wavenumber corresponds to a definite wavelength. For every stored configuration, the hydrodynamic fields are transformed using a normalized FFT. The Fourier amplitudes are then grouped into shells of constant modulus $|\mathbf{k}|$ and averaged over all wavevectors in the same shell (spherical average), yielding the isotropic spectrum
\begin{equation}
S_m(|\mathbf{k}|)=\left\langle |\delta m(\mathbf{k})|^2 \right\rangle .
\end{equation}
To quantify the agreement with equilibrium statistics at each wavelength, we introduce the spectral equilibration ratio
\begin{equation}
\label{eq:er_of_k}
\mathrm{ER}_m(|\mathbf{k}|)=\frac{S_m^{\mathrm{num}}(|\mathbf{k}|)}{S_m^{\mathrm{th}}(|\mathbf{k}|)} ,
\end{equation}
where $S_m^{\mathrm{th}}(|\mathbf{k}|)$ is the theoretical prediction for the same mode. In this representation, departures from unity at small $|\mathbf{k}|$ indicate an imperfect thermalization of long-wavelength fluctuations, whereas discrepancies at large $|\mathbf{k}|$ signal errors at short wavelengths. The resulting $\mathrm{ER}_m(|\mathbf{k}|)$ therefore provides a stringent scale-resolved test of the fluctuation statistics over the set of wavelengths supported by the simulation box.

To better expose model-dependent deviations over the whole range of relaxation times, we introduce a complementary summary based on extremal values. In Figure~\ref{fig:sminmax_comparison} we report, for each $\tau$, the minimum and maximum \emph{spectral} equilibration ratios extracted from the isotropically averaged spectra, namely the extrema of $\mathrm{ER}_m(|\mathbf{k}|)$ over the set of resolved wavenumber shells in the interval $|\mathbf{k}|\in[4,64]$. The results are shown for BGK-FLBM, Reg-FLBM, and GMF-FLBM in four panels corresponding to the modes $m=\rho$, $m=\sum_\alpha \rho u_\alpha$, $m=\sum_{\alpha}\Pi_{\alpha\alpha}$, and $m=\sum_{\alpha<\beta}\Pi_{\alpha\beta}$. This representation condenses the full spectral information into a compact measure of the largest under- and over-shoots with respect to the equilibrium target, and allows a direct comparison of the robustness of the three formulations as $\tau$ is varied.

\begin{figure*}[t]
  \centering
\includegraphics[width=0.95\textwidth]{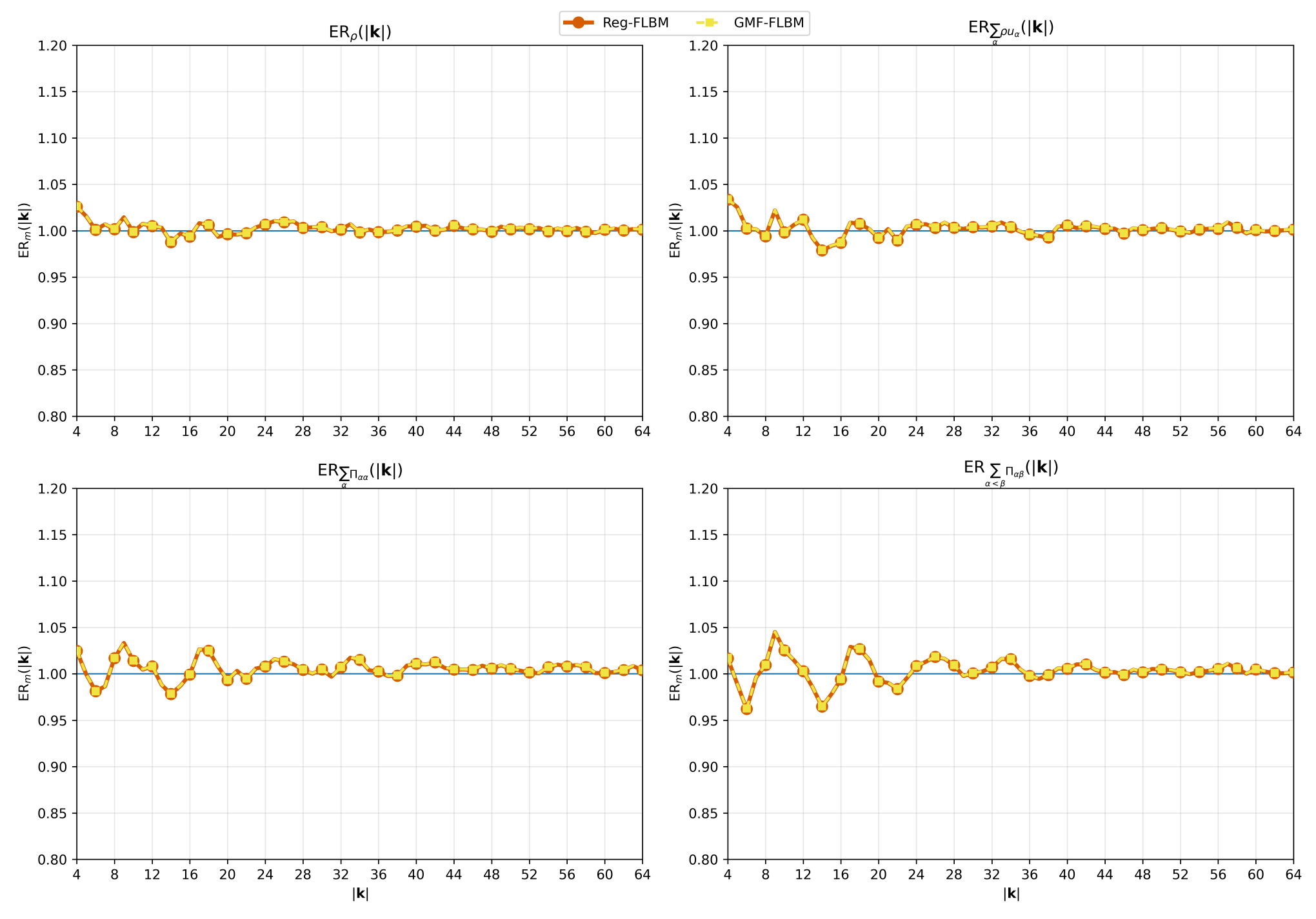}
\caption{Scale-resolved spectral equilibration ratio at $\tau=0.5001$.
Shown is $\mathrm{ER}_m(|\mathbf{k}|)=S_m^{\mathrm{num}}(|\mathbf{k}|)/S_m^{\mathrm{th}}(|\mathbf{k}|)$ for $|\mathbf{k}|\in[4,64]$ and
$m\in\{\rho,\ \sum_\alpha \rho u_\alpha,\ \sum_\alpha \Pi_{\alpha\alpha},\ \sum_{\alpha<\beta}\Pi_{\alpha\beta}\}$, comparing Reg-FLBM and GMF-FLBM.
The two curves overlap across all wavelengths; remaining differences reflect only the collision operator (same noise realization). The line indicates $\mathrm{ER}_m=1$.}
\label{fig:er_k_tau05001}
\end{figure*}

Figure~\ref{fig:sminmax_comparison} highlights the strong sensitivity of the BGK-FLBM to the relaxation time already observed in the literature \cite{lauricella2025regularized,ollila2011fluctuating,bernaschi2009muphy}. 
In the over-relaxed regime close to $\tau\!\to\!1/2^{+}$, and again for very large $\tau$ (under-relaxation), the BGK formulation exhibits pronounced departures from the equilibrium target, with broad extrema of $\mathrm{ER}_m(|\mathbf{k}|)$ across the resolved wavenumber shells. 
These excursions reflect the well-known numerical fragility of single--relaxation-time collisions at very low and very high viscosities \cite{d2002multiple,lallemand2000theory}, and translate into an unreliable scale-by-scale reconstruction of the fluctuation spectrum.
In contrast, GMF-FLBM closely follows the fully regularized Reg-FLBM over the whole range of $\tau$ shown here: the min--max bands remain narrow around $\mathrm{ER}_m=1$ for all four observables, indicating that the ghost-mode filtering preserves fluctuation amplitudes with essentially the same accuracy as the full Hermite regularization, while requiring a markedly simpler collision treatment.

We next focus on the most stringent case, $\tau=0.5001$, where the collision step is maximally over-relaxed and the dynamics is most sensitive to non-hydrodynamic contamination. We therefore examine the full wavenumber dependence of the spectral equilibration ratio defined in Eq.~\eqref{eq:er_of_k}
for the four observables defined above, over the resolved shells $|\mathbf{k}|\in[4,64]$. This scale-resolved view complements the min--max summary and shows how the Reg-FLBM and GMF-FLBM compare across wavelengths, from the longest modes supported by the box to the shortest modes included in the analysis.

In Figure~\ref{fig:er_k_tau05001}, we report the equilibration ratio $\mathrm{ER}_m(|\mathbf{k}|)$ versus the wavenumber magnitude, $|\mathbf{k}|$,  at the most demanding relaxation time, $\tau=0.5001$. Across the whole window $|\mathbf{k}|\in[4,64]$, the GMF-FLBM curves lie essentially on top of the Reg-FLBM ones: the small departures from unity occur at the same wavenumbers and with nearly identical amplitude, indicating that ghost-mode filtering preserves the wavelength-dependent fluctuation content with a comparable accuracy as the full Hermite regularization even in the maximally over-relaxed regime. Importantly, this comparison isolates the effect of the collision operator since the two runs use the same realization of the stochastic forcing (identical random-number sequence). Hence, the residual differences between the curves can only originate from the different collisional treatments rather than from statistical sampling noise.

Importantly, the GMF-FLBM model was able to achieve similar results without the full high-order Hermite reconstruction required by Reg-FLBM.
From an implementation standpoint, GMF-FLBM differs from the BGK update only in replacing the $f_i^{\mathrm{neq}}$ term by the filtered expression introduced above, while leaving the overall structure of the collision step unchanged.
Despite this minimal additional complexity, the ghost-mode filtering removes the dominant source of spurious non-hydrodynamic content and thereby cures the main deficiencies of BGK-FLBM in reproducing equilibrium statistical fluctuations, reaching a quality comparable to Reg-FLBM.

\section{Conclusions and perspectives}
\label{sec:conclusions}

In this work we introduced a ghost-mode filtered fluctuating lattice Boltzmann method (GMF-FLBM) for the D3Q27 lattice, motivated by the observation that, once ghost modes are assigned a unit relaxation, they should not carry deterministic memory and can be reduced to their minimal statistical role. The resulting scheme preserves the hydrodynamic sector while suppressing the propagation of non-hydrodynamic content, replacing it with stochastic contributions consistent with the fluctuation--dissipation theorem.

The numerical tests have shown that ghost-mode filtering substantially improves the robustness of fluctuating LB with respect to the standard single--relaxation-time BGK-FLBM approach over a very wide range of relaxation times, spanning both strongly over-relaxed and strongly under-relaxed regimes. In particular, the equilibration ratios of density, momentum, and stress combinations remain close to unity for GMF-FLBM and are essentially indistinguishable from those of the fully regularized high-order Reg-FLBM, whereas the BGK-FLBM displays pronounced sensitivity to $\tau$ and may even become unstable in the most extreme cases. A scale-resolved analysis in Fourier space confirms that this agreement is not limited to integrated variances. Indeed, the largest deviations in the spectral equilibration ratio stay narrow and centered around the expected fluctuation amplitudes even at the most stringent case $\tau=0.5001$ for all the hydrodynamic observables under investigation.

From a practical standpoint, GMF-FLBM delivers Reg-FLBM-level fluctuation fidelity without requiring the full high-order Hermite reconstruction, and it can be implemented as a minimal modification of the standard BGK update by replacing the non-equilibrium contribution with its filtered expression. This makes ghost-mode filtering an attractive compromise between accuracy and complexity.

Looking ahead, the GMF-FLBM framework could be particularly well suited to soft-matter applications where a fluctuating solvent must remain both faithful and numerically stable across widely separated time and viscosity scales. A natural arena is polymer physics including dilute and semi-dilute solutions, polymer transport, or polymer–colloid suspensions, where one may want to sweep the solvent viscosity from low values (fast hydrodynamic response) to highly viscous conditions representative of concentrated solutions, while still retaining the correct thermal background required for Brownian dynamics and fluctuation-driven phenomena. In this context, the combination of scale-resolved fluctuation accuracy and a safe-thread collision step makes GMF-FLBM an appealing building block for large-scale simulations of polymeric systems, including GPU-accelerated studies where numerical stability and implementation simplicity are essential.

\section*{Data Availability}
The data that support the findings of this study are available from the corresponding author upon reasonable request.

\begin{acknowledgments}
M.L. and A.M. acknowledge funding from the Italian Government through the PRIN (Progetti di Rilevante Interesse Nazionale) Grant (MOBIOS) ID: 2022N4ZNH3 -CUP: F53C24001000006 and computational support of CINECA through the ISCRA B project MIPLAST (HP10BZY7BK). M.L. and A. T. acknowledge the support of the Italian National Group for Mathematical Physics (GNFM-INdAM). M.L., A.T., and S.S. acknowledge the support from the European Research Council under the ERCPoC Grant No. 101187935 (LBFAST).
\end{acknowledgments}

%\bibliographystyle{alpha}
%\bibliography{sample}

\newpage
%\bibliographystyle{aipnum4-1} 
%\bibliography{bibliography}

\begin{thebibliography}{61}%
\makeatletter
\providecommand \@ifxundefined [1]{%
 \@ifx{#1\undefined}
}%
\providecommand \@ifnum [1]{%
 \ifnum #1\expandafter \@firstoftwo
 \else \expandafter \@secondoftwo
 \fi
}%
\providecommand \@ifx [1]{%
 \ifx #1\expandafter \@firstoftwo
 \else \expandafter \@secondoftwo
 \fi
}%
\providecommand \natexlab [1]{#1}%
\providecommand \enquote  [1]{``#1''}%
\providecommand \bibnamefont  [1]{#1}%
\providecommand \bibfnamefont [1]{#1}%
\providecommand \citenamefont [1]{#1}%
\providecommand \href@noop [0]{\@secondoftwo}%
\providecommand \href [0]{\begingroup \@sanitize@url \@href}%
\providecommand \@href[1]{\@@startlink{#1}\@@href}%
\providecommand \@@href[1]{\endgroup#1\@@endlink}%
\providecommand \@sanitize@url [0]{\catcode `\\12\catcode `\$12\catcode
  `\&12\catcode `\#12\catcode `\^12\catcode `\_12\catcode `\%12\relax}%
\providecommand \@@startlink[1]{}%
\providecommand \@@endlink[0]{}%
\providecommand \url  [0]{\begingroup\@sanitize@url \@url }%
\providecommand \@url [1]{\endgroup\@href {#1}{\urlprefix }}%
\providecommand \urlprefix  [0]{URL }%
\providecommand \Eprint [0]{\href }%
\providecommand \doibase [0]{http://dx.doi.org/}%
\providecommand \selectlanguage [0]{\@gobble}%
\providecommand \bibinfo  [0]{\@secondoftwo}%
\providecommand \bibfield  [0]{\@secondoftwo}%
\providecommand \translation [1]{[#1]}%
\providecommand \BibitemOpen [0]{}%
\providecommand \bibitemStop [0]{}%
\providecommand \bibitemNoStop [0]{.\EOS\space}%
\providecommand \EOS [0]{\spacefactor3000\relax}%
\providecommand \BibitemShut  [1]{\csname bibitem#1\endcsname}%
\let\auto@bib@innerbib\@empty
%</preamble>
\bibitem [{\citenamefont {Tiribocchi}\ \emph {et~al.}(2025)\citenamefont
  {Tiribocchi}, \citenamefont {Durve}, \citenamefont {Lauricella},
  \citenamefont {Montessori}, \citenamefont {Tucny},\ and\ \citenamefont
  {Succi}}]{tiribocchi2025lattice}%
  \BibitemOpen
  \bibfield  {author} {\bibinfo {author} {\bibfnamefont {A.}~\bibnamefont
  {Tiribocchi}}, \bibinfo {author} {\bibfnamefont {M.}~\bibnamefont {Durve}},
  \bibinfo {author} {\bibfnamefont {M.}~\bibnamefont {Lauricella}}, \bibinfo
  {author} {\bibfnamefont {A.}~\bibnamefont {Montessori}}, \bibinfo {author}
  {\bibfnamefont {J.-M.}\ \bibnamefont {Tucny}}, \ and\ \bibinfo {author}
  {\bibfnamefont {S.}~\bibnamefont {Succi}},\ }\href@noop {} {\bibfield
  {journal} {\bibinfo  {journal} {Physics Reports}\ }\textbf {\bibinfo {volume}
  {1105}},\ \bibinfo {pages} {1} (\bibinfo {year} {2025})}\BibitemShut
  {NoStop}%
\bibitem [{\citenamefont {Succi}(2018)}]{succi2018lattice}%
  \BibitemOpen
  \bibfield  {author} {\bibinfo {author} {\bibfnamefont {S.}~\bibnamefont
  {Succi}},\ }\href@noop {} {\emph {\bibinfo {title} {The lattice Boltzmann
  equation: for complex states of flowing matter}}}\ (\bibinfo  {publisher}
  {Oxford university press},\ \bibinfo {year} {2018})\BibitemShut {NoStop}%
\bibitem [{\citenamefont {Kr{\"u}ger}\ \emph {et~al.}(2017)\citenamefont
  {Kr{\"u}ger}, \citenamefont {Kusumaatmaja}, \citenamefont {Kuzmin},
  \citenamefont {Shardt}, \citenamefont {Silva},\ and\ \citenamefont
  {Viggen}}]{kruger2017lattice}%
  \BibitemOpen
  \bibfield  {author} {\bibinfo {author} {\bibfnamefont {T.}~\bibnamefont
  {Kr{\"u}ger}}, \bibinfo {author} {\bibfnamefont {H.}~\bibnamefont
  {Kusumaatmaja}}, \bibinfo {author} {\bibfnamefont {A.}~\bibnamefont
  {Kuzmin}}, \bibinfo {author} {\bibfnamefont {O.}~\bibnamefont {Shardt}},
  \bibinfo {author} {\bibfnamefont {G.}~\bibnamefont {Silva}}, \ and\ \bibinfo
  {author} {\bibfnamefont {E.~M.}\ \bibnamefont {Viggen}},\ }\href@noop {}
  {\bibfield  {journal} {\bibinfo  {journal} {Springer International
  Publishing}\ }\textbf {\bibinfo {volume} {10}},\ \bibinfo {pages} {4}
  (\bibinfo {year} {2017})}\BibitemShut {NoStop}%
\bibitem [{\citenamefont {Sukop}\ and\ \citenamefont
  {Thorne~Jr}(2006)}]{sukop2006lattice}%
  \BibitemOpen
  \bibfield  {author} {\bibinfo {author} {\bibfnamefont {M.~C.}\ \bibnamefont
  {Sukop}}\ and\ \bibinfo {author} {\bibfnamefont {D.~T.}\ \bibnamefont
  {Thorne~Jr}},\ }\href@noop {} {\emph {\bibinfo {title} {Lattice Boltzmann
  modeling: an introduction for geoscientists and engineers}}}\ (\bibinfo
  {publisher} {Springer},\ \bibinfo {year} {2006})\BibitemShut {NoStop}%
\bibitem [{\citenamefont {Boek}\ and\ \citenamefont
  {Venturoli}(2010)}]{boek2010lattice}%
  \BibitemOpen
  \bibfield  {author} {\bibinfo {author} {\bibfnamefont {E.~S.}\ \bibnamefont
  {Boek}}\ and\ \bibinfo {author} {\bibfnamefont {M.}~\bibnamefont
  {Venturoli}},\ }\href@noop {} {\bibfield  {journal} {\bibinfo  {journal}
  {Computers \& Mathematics with Applications}\ }\textbf {\bibinfo {volume}
  {59}},\ \bibinfo {pages} {2305} (\bibinfo {year} {2010})}\BibitemShut
  {NoStop}%
\bibitem [{\citenamefont {Guo}\ and\ \citenamefont
  {Zhao}(2002)}]{guo2002lattice}%
  \BibitemOpen
  \bibfield  {author} {\bibinfo {author} {\bibfnamefont {Z.}~\bibnamefont
  {Guo}}\ and\ \bibinfo {author} {\bibfnamefont {T.}~\bibnamefont {Zhao}},\
  }\href@noop {} {\bibfield  {journal} {\bibinfo  {journal} {Physical review
  E}\ }\textbf {\bibinfo {volume} {66}},\ \bibinfo {pages} {036304} (\bibinfo
  {year} {2002})}\BibitemShut {NoStop}%
\bibitem [{\citenamefont {Cali}\ \emph {et~al.}(1992)\citenamefont {Cali},
  \citenamefont {Succi}, \citenamefont {Cancelliere}, \citenamefont {Benzi},\
  and\ \citenamefont {Gramignani}}]{cali1992diffusion}%
  \BibitemOpen
  \bibfield  {author} {\bibinfo {author} {\bibfnamefont {A.}~\bibnamefont
  {Cali}}, \bibinfo {author} {\bibfnamefont {S.}~\bibnamefont {Succi}},
  \bibinfo {author} {\bibfnamefont {A.}~\bibnamefont {Cancelliere}}, \bibinfo
  {author} {\bibfnamefont {R.}~\bibnamefont {Benzi}}, \ and\ \bibinfo {author}
  {\bibfnamefont {M.}~\bibnamefont {Gramignani}},\ }\href@noop {} {\bibfield
  {journal} {\bibinfo  {journal} {Physical Review A}\ }\textbf {\bibinfo
  {volume} {45}},\ \bibinfo {pages} {5771} (\bibinfo {year}
  {1992})}\BibitemShut {NoStop}%
\bibitem [{\citenamefont {Montessori}, \citenamefont {Hegele},\ and\
  \citenamefont {Lauricella}(2025)}]{montessori2025thread}%
  \BibitemOpen
  \bibfield  {author} {\bibinfo {author} {\bibfnamefont {A.}~\bibnamefont
  {Montessori}}, \bibinfo {author} {\bibfnamefont {L.~A.}\ \bibnamefont
  {Hegele}}, \ and\ \bibinfo {author} {\bibfnamefont {M.}~\bibnamefont
  {Lauricella}},\ }\href@noop {} {\bibfield  {journal} {\bibinfo  {journal}
  {AIAA Journal}\ }\textbf {\bibinfo {volume} {63}},\ \bibinfo {pages} {1005}
  (\bibinfo {year} {2025})}\BibitemShut {NoStop}%
\bibitem [{\citenamefont {Zhang}\ \emph {et~al.}(2024)\citenamefont {Zhang},
  \citenamefont {Li}, \citenamefont {Wang},\ and\ \citenamefont
  {Shu}}]{zhang2024improved}%
  \BibitemOpen
  \bibfield  {author} {\bibinfo {author} {\bibfnamefont {D.}~\bibnamefont
  {Zhang}}, \bibinfo {author} {\bibfnamefont {Y.}~\bibnamefont {Li}}, \bibinfo
  {author} {\bibfnamefont {Y.}~\bibnamefont {Wang}}, \ and\ \bibinfo {author}
  {\bibfnamefont {C.}~\bibnamefont {Shu}},\ }\href@noop {} {\bibfield
  {journal} {\bibinfo  {journal} {Physics of Fluids}\ }\textbf {\bibinfo
  {volume} {36}} (\bibinfo {year} {2024})}\BibitemShut {NoStop}%
\bibitem [{\citenamefont {Tiribocchi}\ \emph {et~al.}(2020)\citenamefont
  {Tiribocchi}, \citenamefont {Montessori}, \citenamefont {Aime}, \citenamefont
  {Milani}, \citenamefont {Lauricella}, \citenamefont {Succi},\ and\
  \citenamefont {Weitz}}]{tiribocchi2020novel}%
  \BibitemOpen
  \bibfield  {author} {\bibinfo {author} {\bibfnamefont {A.}~\bibnamefont
  {Tiribocchi}}, \bibinfo {author} {\bibfnamefont {A.}~\bibnamefont
  {Montessori}}, \bibinfo {author} {\bibfnamefont {S.}~\bibnamefont {Aime}},
  \bibinfo {author} {\bibfnamefont {M.}~\bibnamefont {Milani}}, \bibinfo
  {author} {\bibfnamefont {M.}~\bibnamefont {Lauricella}}, \bibinfo {author}
  {\bibfnamefont {S.}~\bibnamefont {Succi}}, \ and\ \bibinfo {author}
  {\bibfnamefont {D.}~\bibnamefont {Weitz}},\ }\href@noop {} {\bibfield
  {journal} {\bibinfo  {journal} {Physics of Fluids}\ }\textbf {\bibinfo
  {volume} {32}} (\bibinfo {year} {2020})}\BibitemShut {NoStop}%
\bibitem [{\citenamefont {Chiappini}\ \emph {et~al.}(2019)\citenamefont
  {Chiappini}, \citenamefont {Sbragaglia}, \citenamefont {Xue},\ and\
  \citenamefont {Falcucci}}]{chiappini2019hydrodynamic}%
  \BibitemOpen
  \bibfield  {author} {\bibinfo {author} {\bibfnamefont {D.}~\bibnamefont
  {Chiappini}}, \bibinfo {author} {\bibfnamefont {M.}~\bibnamefont
  {Sbragaglia}}, \bibinfo {author} {\bibfnamefont {X.}~\bibnamefont {Xue}}, \
  and\ \bibinfo {author} {\bibfnamefont {G.}~\bibnamefont {Falcucci}},\
  }\href@noop {} {\bibfield  {journal} {\bibinfo  {journal} {Physical Review
  E}\ }\textbf {\bibinfo {volume} {99}},\ \bibinfo {pages} {053305} (\bibinfo
  {year} {2019})}\BibitemShut {NoStop}%
\bibitem [{\citenamefont {Montessori}, \citenamefont {Lauricella},\ and\
  \citenamefont {Succi}(2019)}]{montessori2019mesoscale}%
  \BibitemOpen
  \bibfield  {author} {\bibinfo {author} {\bibfnamefont {A.}~\bibnamefont
  {Montessori}}, \bibinfo {author} {\bibfnamefont {M.}~\bibnamefont
  {Lauricella}}, \ and\ \bibinfo {author} {\bibfnamefont {S.}~\bibnamefont
  {Succi}},\ }\href@noop {} {\bibfield  {journal} {\bibinfo  {journal}
  {Philosophical Transactions of the Royal Society A}\ }\textbf {\bibinfo
  {volume} {377}},\ \bibinfo {pages} {20180149} (\bibinfo {year}
  {2019})}\BibitemShut {NoStop}%
\bibitem [{\citenamefont {Liu}, \citenamefont {Valocchi},\ and\ \citenamefont
  {Kang}(2012)}]{liu2012three}%
  \BibitemOpen
  \bibfield  {author} {\bibinfo {author} {\bibfnamefont {H.}~\bibnamefont
  {Liu}}, \bibinfo {author} {\bibfnamefont {A.~J.}\ \bibnamefont {Valocchi}}, \
  and\ \bibinfo {author} {\bibfnamefont {Q.}~\bibnamefont {Kang}},\ }\href@noop
  {} {\bibfield  {journal} {\bibinfo  {journal} {Physical Review
  E—Statistical, Nonlinear, and Soft Matter Physics}\ }\textbf {\bibinfo
  {volume} {85}},\ \bibinfo {pages} {046309} (\bibinfo {year}
  {2012})}\BibitemShut {NoStop}%
\bibitem [{\citenamefont {Shan}\ and\ \citenamefont
  {Chen}(1993)}]{shan1993lattice}%
  \BibitemOpen
  \bibfield  {author} {\bibinfo {author} {\bibfnamefont {X.}~\bibnamefont
  {Shan}}\ and\ \bibinfo {author} {\bibfnamefont {H.}~\bibnamefont {Chen}},\
  }\href@noop {} {\bibfield  {journal} {\bibinfo  {journal} {Physical review
  E}\ }\textbf {\bibinfo {volume} {47}},\ \bibinfo {pages} {1815} (\bibinfo
  {year} {1993})}\BibitemShut {NoStop}%
\bibitem [{\citenamefont {Xiong}\ \emph {et~al.}(2025)\citenamefont {Xiong},
  \citenamefont {Wang}, \citenamefont {Huang},\ and\ \citenamefont
  {Luo}}]{xiong2025thermodynamically}%
  \BibitemOpen
  \bibfield  {author} {\bibinfo {author} {\bibfnamefont {F.}~\bibnamefont
  {Xiong}}, \bibinfo {author} {\bibfnamefont {L.}~\bibnamefont {Wang}},
  \bibinfo {author} {\bibfnamefont {J.}~\bibnamefont {Huang}}, \ and\ \bibinfo
  {author} {\bibfnamefont {K.}~\bibnamefont {Luo}},\ }\href@noop {} {\bibfield
  {journal} {\bibinfo  {journal} {Journal of Scientific Computing}\ }\textbf
  {\bibinfo {volume} {103}},\ \bibinfo {pages} {1} (\bibinfo {year}
  {2025})}\BibitemShut {NoStop}%
\bibitem [{\citenamefont {Liu}\ \emph {et~al.}(2024)\citenamefont {Liu},
  \citenamefont {Chai}, \citenamefont {Shi},\ and\ \citenamefont
  {Yuan}}]{liu2024consistent}%
  \BibitemOpen
  \bibfield  {author} {\bibinfo {author} {\bibfnamefont {X.}~\bibnamefont
  {Liu}}, \bibinfo {author} {\bibfnamefont {Z.}~\bibnamefont {Chai}}, \bibinfo
  {author} {\bibfnamefont {B.}~\bibnamefont {Shi}}, \ and\ \bibinfo {author}
  {\bibfnamefont {X.}~\bibnamefont {Yuan}},\ }\href@noop {} {\bibfield
  {journal} {\bibinfo  {journal} {Physica D: Nonlinear Phenomena}\ }\textbf
  {\bibinfo {volume} {468}},\ \bibinfo {pages} {134294} (\bibinfo {year}
  {2024})}\BibitemShut {NoStop}%
\bibitem [{\citenamefont {Wang}\ \emph {et~al.}(2021)\citenamefont {Wang},
  \citenamefont {Wei}, \citenamefont {Li}, \citenamefont {Chai},\ and\
  \citenamefont {Shi}}]{wang2021lattice}%
  \BibitemOpen
  \bibfield  {author} {\bibinfo {author} {\bibfnamefont {L.}~\bibnamefont
  {Wang}}, \bibinfo {author} {\bibfnamefont {Z.}~\bibnamefont {Wei}}, \bibinfo
  {author} {\bibfnamefont {T.}~\bibnamefont {Li}}, \bibinfo {author}
  {\bibfnamefont {Z.}~\bibnamefont {Chai}}, \ and\ \bibinfo {author}
  {\bibfnamefont {B.}~\bibnamefont {Shi}},\ }\href@noop {} {\bibfield
  {journal} {\bibinfo  {journal} {Applied Mathematical Modelling}\ }\textbf
  {\bibinfo {volume} {95}},\ \bibinfo {pages} {361} (\bibinfo {year}
  {2021})}\BibitemShut {NoStop}%
\bibitem [{\citenamefont {Lauricella}\ \emph {et~al.}(2018)\citenamefont
  {Lauricella}, \citenamefont {Melchionna}, \citenamefont {Montessori},
  \citenamefont {Pisignano}, \citenamefont {Pontrelli},\ and\ \citenamefont
  {Succi}}]{lauricella2018entropic}%
  \BibitemOpen
  \bibfield  {author} {\bibinfo {author} {\bibfnamefont {M.}~\bibnamefont
  {Lauricella}}, \bibinfo {author} {\bibfnamefont {S.}~\bibnamefont
  {Melchionna}}, \bibinfo {author} {\bibfnamefont {A.}~\bibnamefont
  {Montessori}}, \bibinfo {author} {\bibfnamefont {D.}~\bibnamefont
  {Pisignano}}, \bibinfo {author} {\bibfnamefont {G.}~\bibnamefont
  {Pontrelli}}, \ and\ \bibinfo {author} {\bibfnamefont {S.}~\bibnamefont
  {Succi}},\ }\href@noop {} {\bibfield  {journal} {\bibinfo  {journal}
  {Physical Review E}\ }\textbf {\bibinfo {volume} {97}},\ \bibinfo {pages}
  {033308} (\bibinfo {year} {2018})}\BibitemShut {NoStop}%
\bibitem [{\citenamefont {Kupershtokh}\ and\ \citenamefont
  {Medvedev}(2006)}]{kupershtokh2006lattice}%
  \BibitemOpen
  \bibfield  {author} {\bibinfo {author} {\bibfnamefont {A.}~\bibnamefont
  {Kupershtokh}}\ and\ \bibinfo {author} {\bibfnamefont {D.}~\bibnamefont
  {Medvedev}},\ }\href@noop {} {\bibfield  {journal} {\bibinfo  {journal}
  {Journal of electrostatics}\ }\textbf {\bibinfo {volume} {64}},\ \bibinfo
  {pages} {581} (\bibinfo {year} {2006})}\BibitemShut {NoStop}%
\bibitem [{\citenamefont {Guglietta}\ \emph {et~al.}(2023)\citenamefont
  {Guglietta}, \citenamefont {Pelusi}, \citenamefont {Sega}, \citenamefont
  {Aouane},\ and\ \citenamefont {Harting}}]{guglietta2023suspensions}%
  \BibitemOpen
  \bibfield  {author} {\bibinfo {author} {\bibfnamefont {F.}~\bibnamefont
  {Guglietta}}, \bibinfo {author} {\bibfnamefont {F.}~\bibnamefont {Pelusi}},
  \bibinfo {author} {\bibfnamefont {M.}~\bibnamefont {Sega}}, \bibinfo {author}
  {\bibfnamefont {O.}~\bibnamefont {Aouane}}, \ and\ \bibinfo {author}
  {\bibfnamefont {J.}~\bibnamefont {Harting}},\ }\href@noop {} {\bibfield
  {journal} {\bibinfo  {journal} {Journal of Fluid Mechanics}\ }\textbf
  {\bibinfo {volume} {971}},\ \bibinfo {pages} {A13} (\bibinfo {year}
  {2023})}\BibitemShut {NoStop}%
\bibitem [{\citenamefont {Yang}\ \emph {et~al.}(2022)\citenamefont {Yang},
  \citenamefont {Sega}, \citenamefont {Leimbach}, \citenamefont {Kolb},
  \citenamefont {Karl},\ and\ \citenamefont {Harting}}]{yang2022capillary}%
  \BibitemOpen
  \bibfield  {author} {\bibinfo {author} {\bibfnamefont {L.}~\bibnamefont
  {Yang}}, \bibinfo {author} {\bibfnamefont {M.}~\bibnamefont {Sega}}, \bibinfo
  {author} {\bibfnamefont {S.}~\bibnamefont {Leimbach}}, \bibinfo {author}
  {\bibfnamefont {S.}~\bibnamefont {Kolb}}, \bibinfo {author} {\bibfnamefont
  {J.}~\bibnamefont {Karl}}, \ and\ \bibinfo {author} {\bibfnamefont
  {J.}~\bibnamefont {Harting}},\ }\href@noop {} {\bibfield  {journal} {\bibinfo
   {journal} {Industrial \& engineering chemistry research}\ }\textbf {\bibinfo
  {volume} {61}},\ \bibinfo {pages} {1863} (\bibinfo {year}
  {2022})}\BibitemShut {NoStop}%
\bibitem [{\citenamefont {Bonaccorso}\ \emph {et~al.}(2020)\citenamefont
  {Bonaccorso}, \citenamefont {Montessori}, \citenamefont {Tiribocchi},
  \citenamefont {Amati}, \citenamefont {Bernaschi}, \citenamefont
  {Lauricella},\ and\ \citenamefont {Succi}}]{bonaccorso2020lbsoft}%
  \BibitemOpen
  \bibfield  {author} {\bibinfo {author} {\bibfnamefont {F.}~\bibnamefont
  {Bonaccorso}}, \bibinfo {author} {\bibfnamefont {A.}~\bibnamefont
  {Montessori}}, \bibinfo {author} {\bibfnamefont {A.}~\bibnamefont
  {Tiribocchi}}, \bibinfo {author} {\bibfnamefont {G.}~\bibnamefont {Amati}},
  \bibinfo {author} {\bibfnamefont {M.}~\bibnamefont {Bernaschi}}, \bibinfo
  {author} {\bibfnamefont {M.}~\bibnamefont {Lauricella}}, \ and\ \bibinfo
  {author} {\bibfnamefont {S.}~\bibnamefont {Succi}},\ }\href@noop {}
  {\bibfield  {journal} {\bibinfo  {journal} {Computer Physics Communications}\
  }\textbf {\bibinfo {volume} {256}},\ \bibinfo {pages} {107455} (\bibinfo
  {year} {2020})}\BibitemShut {NoStop}%
\bibitem [{\citenamefont {Harting}\ \emph {et~al.}(2014)\citenamefont
  {Harting}, \citenamefont {Frijters}, \citenamefont {Ramaioli}, \citenamefont
  {Robinson}, \citenamefont {Wolf},\ and\ \citenamefont
  {Luding}}]{harting2014recent}%
  \BibitemOpen
  \bibfield  {author} {\bibinfo {author} {\bibfnamefont {J.}~\bibnamefont
  {Harting}}, \bibinfo {author} {\bibfnamefont {S.}~\bibnamefont {Frijters}},
  \bibinfo {author} {\bibfnamefont {M.}~\bibnamefont {Ramaioli}}, \bibinfo
  {author} {\bibfnamefont {M.}~\bibnamefont {Robinson}}, \bibinfo {author}
  {\bibfnamefont {D.~E.}\ \bibnamefont {Wolf}}, \ and\ \bibinfo {author}
  {\bibfnamefont {S.}~\bibnamefont {Luding}},\ }\href@noop {} {\bibfield
  {journal} {\bibinfo  {journal} {The European Physical Journal Special
  Topics}\ }\textbf {\bibinfo {volume} {223}},\ \bibinfo {pages} {2253}
  (\bibinfo {year} {2014})}\BibitemShut {NoStop}%
\bibitem [{\citenamefont {Ladd}\ and\ \citenamefont
  {Verberg}(2001)}]{ladd2001lattice}%
  \BibitemOpen
  \bibfield  {author} {\bibinfo {author} {\bibfnamefont {A.~J.}\ \bibnamefont
  {Ladd}}\ and\ \bibinfo {author} {\bibfnamefont {R.}~\bibnamefont {Verberg}},\
  }\href@noop {} {\bibfield  {journal} {\bibinfo  {journal} {Journal of
  statistical physics}\ }\textbf {\bibinfo {volume} {104}},\ \bibinfo {pages}
  {1191} (\bibinfo {year} {2001})}\BibitemShut {NoStop}%
\bibitem [{\citenamefont {Monteferrante}\ \emph {et~al.}(2021)\citenamefont
  {Monteferrante}, \citenamefont {Montessori}, \citenamefont {Succi},
  \citenamefont {Pisignano},\ and\ \citenamefont
  {Lauricella}}]{monteferrante2021lattice}%
  \BibitemOpen
  \bibfield  {author} {\bibinfo {author} {\bibfnamefont {M.}~\bibnamefont
  {Monteferrante}}, \bibinfo {author} {\bibfnamefont {A.}~\bibnamefont
  {Montessori}}, \bibinfo {author} {\bibfnamefont {S.}~\bibnamefont {Succi}},
  \bibinfo {author} {\bibfnamefont {D.}~\bibnamefont {Pisignano}}, \ and\
  \bibinfo {author} {\bibfnamefont {M.}~\bibnamefont {Lauricella}},\
  }\href@noop {} {\bibfield  {journal} {\bibinfo  {journal} {Physics of
  Fluids}\ }\textbf {\bibinfo {volume} {33}} (\bibinfo {year}
  {2021})}\BibitemShut {NoStop}%
\bibitem [{\citenamefont {Malaspinas}, \citenamefont {Fi{\'e}tier},\ and\
  \citenamefont {Deville}(2010)}]{malaspinas2010lattice}%
  \BibitemOpen
  \bibfield  {author} {\bibinfo {author} {\bibfnamefont {O.}~\bibnamefont
  {Malaspinas}}, \bibinfo {author} {\bibfnamefont {N.}~\bibnamefont
  {Fi{\'e}tier}}, \ and\ \bibinfo {author} {\bibfnamefont {M.}~\bibnamefont
  {Deville}},\ }\href@noop {} {\bibfield  {journal} {\bibinfo  {journal}
  {Journal of Non-Newtonian Fluid Mechanics}\ }\textbf {\bibinfo {volume}
  {165}},\ \bibinfo {pages} {1637} (\bibinfo {year} {2010})}\BibitemShut
  {NoStop}%
\bibitem [{\citenamefont {Berk~Usta}, \citenamefont {Ladd},\ and\ \citenamefont
  {Butler}(2005)}]{berk2005lattice}%
  \BibitemOpen
  \bibfield  {author} {\bibinfo {author} {\bibfnamefont {O.}~\bibnamefont
  {Berk~Usta}}, \bibinfo {author} {\bibfnamefont {A.~J.}\ \bibnamefont {Ladd}},
  \ and\ \bibinfo {author} {\bibfnamefont {J.~E.}\ \bibnamefont {Butler}},\
  }\href@noop {} {\bibfield  {journal} {\bibinfo  {journal} {The Journal of
  chemical physics}\ }\textbf {\bibinfo {volume} {122}} (\bibinfo {year}
  {2005})}\BibitemShut {NoStop}%
\bibitem [{\citenamefont {Ahlrichs}\ and\ \citenamefont
  {D{\"u}nweg}(1999)}]{ahlrichs1999simulation}%
  \BibitemOpen
  \bibfield  {author} {\bibinfo {author} {\bibfnamefont {P.}~\bibnamefont
  {Ahlrichs}}\ and\ \bibinfo {author} {\bibfnamefont {B.}~\bibnamefont
  {D{\"u}nweg}},\ }\href@noop {} {\bibfield  {journal} {\bibinfo  {journal}
  {The Journal of chemical physics}\ }\textbf {\bibinfo {volume} {111}},\
  \bibinfo {pages} {8225} (\bibinfo {year} {1999})}\BibitemShut {NoStop}%
\bibitem [{\citenamefont {Ahlrichs}\ and\ \citenamefont
  {D{\"u}nweg}(1998)}]{ahlrichs1998lattice}%
  \BibitemOpen
  \bibfield  {author} {\bibinfo {author} {\bibfnamefont {P.}~\bibnamefont
  {Ahlrichs}}\ and\ \bibinfo {author} {\bibfnamefont {B.}~\bibnamefont
  {D{\"u}nweg}},\ }\href@noop {} {\bibfield  {journal} {\bibinfo  {journal}
  {International Journal of Modern Physics C}\ }\textbf {\bibinfo {volume}
  {9}},\ \bibinfo {pages} {1429} (\bibinfo {year} {1998})}\BibitemShut
  {NoStop}%
\bibitem [{\citenamefont {Sawant}, \citenamefont {Dorschner},\ and\
  \citenamefont {Karlin}(2021)}]{sawant2021lattice}%
  \BibitemOpen
  \bibfield  {author} {\bibinfo {author} {\bibfnamefont {N.}~\bibnamefont
  {Sawant}}, \bibinfo {author} {\bibfnamefont {B.}~\bibnamefont {Dorschner}}, \
  and\ \bibinfo {author} {\bibfnamefont {I.~V.}\ \bibnamefont {Karlin}},\
  }\href@noop {} {\bibfield  {journal} {\bibinfo  {journal} {Philosophical
  Transactions of the Royal Society A}\ }\textbf {\bibinfo {volume} {379}},\
  \bibinfo {pages} {20200402} (\bibinfo {year} {2021})}\BibitemShut {NoStop}%
\bibitem [{\citenamefont {Lin}\ \emph {et~al.}(2017)\citenamefont {Lin},
  \citenamefont {Luo}, \citenamefont {Fei},\ and\ \citenamefont
  {Succi}}]{lin2017multi}%
  \BibitemOpen
  \bibfield  {author} {\bibinfo {author} {\bibfnamefont {C.}~\bibnamefont
  {Lin}}, \bibinfo {author} {\bibfnamefont {K.~H.}\ \bibnamefont {Luo}},
  \bibinfo {author} {\bibfnamefont {L.}~\bibnamefont {Fei}}, \ and\ \bibinfo
  {author} {\bibfnamefont {S.}~\bibnamefont {Succi}},\ }\href@noop {}
  {\bibfield  {journal} {\bibinfo  {journal} {Scientific reports}\ }\textbf
  {\bibinfo {volume} {7}},\ \bibinfo {pages} {14580} (\bibinfo {year}
  {2017})}\BibitemShut {NoStop}%
\bibitem [{\citenamefont {Tiribocchi}\ \emph {et~al.}(2023)\citenamefont
  {Tiribocchi}, \citenamefont {Durve}, \citenamefont {Lauricella},
  \citenamefont {Montessori}, \citenamefont {Marenduzzo},\ and\ \citenamefont
  {Succi}}]{tiribocchi2023crucial}%
  \BibitemOpen
  \bibfield  {author} {\bibinfo {author} {\bibfnamefont {A.}~\bibnamefont
  {Tiribocchi}}, \bibinfo {author} {\bibfnamefont {M.}~\bibnamefont {Durve}},
  \bibinfo {author} {\bibfnamefont {M.}~\bibnamefont {Lauricella}}, \bibinfo
  {author} {\bibfnamefont {A.}~\bibnamefont {Montessori}}, \bibinfo {author}
  {\bibfnamefont {D.}~\bibnamefont {Marenduzzo}}, \ and\ \bibinfo {author}
  {\bibfnamefont {S.}~\bibnamefont {Succi}},\ }\href@noop {} {\bibfield
  {journal} {\bibinfo  {journal} {Nature Communications}\ }\textbf {\bibinfo
  {volume} {14}},\ \bibinfo {pages} {1096} (\bibinfo {year}
  {2023})}\BibitemShut {NoStop}%
\bibitem [{\citenamefont {Carenza}\ \emph {et~al.}(2020)\citenamefont
  {Carenza}, \citenamefont {Gonnella}, \citenamefont {Marenduzzo},\ and\
  \citenamefont {Negro}}]{carenza2020chaotic}%
  \BibitemOpen
  \bibfield  {author} {\bibinfo {author} {\bibfnamefont {L.~N.}\ \bibnamefont
  {Carenza}}, \bibinfo {author} {\bibfnamefont {G.}~\bibnamefont {Gonnella}},
  \bibinfo {author} {\bibfnamefont {D.}~\bibnamefont {Marenduzzo}}, \ and\
  \bibinfo {author} {\bibfnamefont {G.}~\bibnamefont {Negro}},\ }\href@noop {}
  {\bibfield  {journal} {\bibinfo  {journal} {Physica A: Statistical Mechanics
  and its Applications}\ }\textbf {\bibinfo {volume} {559}},\ \bibinfo {pages}
  {125025} (\bibinfo {year} {2020})}\BibitemShut {NoStop}%
\bibitem [{\citenamefont {Doostmohammadi}\ \emph {et~al.}(2016)\citenamefont
  {Doostmohammadi}, \citenamefont {Adamer}, \citenamefont {Thampi},\ and\
  \citenamefont {Yeomans}}]{doostmohammadi2016stabilization}%
  \BibitemOpen
  \bibfield  {author} {\bibinfo {author} {\bibfnamefont {A.}~\bibnamefont
  {Doostmohammadi}}, \bibinfo {author} {\bibfnamefont {M.~F.}\ \bibnamefont
  {Adamer}}, \bibinfo {author} {\bibfnamefont {S.~P.}\ \bibnamefont {Thampi}},
  \ and\ \bibinfo {author} {\bibfnamefont {J.~M.}\ \bibnamefont {Yeomans}},\
  }\href@noop {} {\bibfield  {journal} {\bibinfo  {journal} {Nature
  communications}\ }\textbf {\bibinfo {volume} {7}},\ \bibinfo {pages} {10557}
  (\bibinfo {year} {2016})}\BibitemShut {NoStop}%
\bibitem [{\citenamefont {Marenduzzo}\ \emph {et~al.}(2007)\citenamefont
  {Marenduzzo}, \citenamefont {Orlandini}, \citenamefont {Cates},\ and\
  \citenamefont {Yeomans}}]{marenduzzo2007steady}%
  \BibitemOpen
  \bibfield  {author} {\bibinfo {author} {\bibfnamefont {D.}~\bibnamefont
  {Marenduzzo}}, \bibinfo {author} {\bibfnamefont {E.}~\bibnamefont
  {Orlandini}}, \bibinfo {author} {\bibfnamefont {M.}~\bibnamefont {Cates}}, \
  and\ \bibinfo {author} {\bibfnamefont {J.}~\bibnamefont {Yeomans}},\
  }\href@noop {} {\bibfield  {journal} {\bibinfo  {journal} {Physical Review
  E—Statistical, Nonlinear, and Soft Matter Physics}\ }\textbf {\bibinfo
  {volume} {76}},\ \bibinfo {pages} {031921} (\bibinfo {year}
  {2007})}\BibitemShut {NoStop}%
\bibitem [{\citenamefont {Adhikari}\ \emph {et~al.}(2005)\citenamefont
  {Adhikari}, \citenamefont {Stratford}, \citenamefont {Cates},\ and\
  \citenamefont {Wagner}}]{adhikari2005fluctuating}%
  \BibitemOpen
  \bibfield  {author} {\bibinfo {author} {\bibfnamefont {R.}~\bibnamefont
  {Adhikari}}, \bibinfo {author} {\bibfnamefont {K.}~\bibnamefont {Stratford}},
  \bibinfo {author} {\bibfnamefont {M.}~\bibnamefont {Cates}}, \ and\ \bibinfo
  {author} {\bibfnamefont {A.}~\bibnamefont {Wagner}},\ }\href@noop {}
  {\bibfield  {journal} {\bibinfo  {journal} {Europhysics Letters}\ }\textbf
  {\bibinfo {volume} {71}},\ \bibinfo {pages} {473} (\bibinfo {year}
  {2005})}\BibitemShut {NoStop}%
\bibitem [{\citenamefont {D{\"u}nweg}\ and\ \citenamefont
  {Kremer}(1993)}]{dunweg1993molecular}%
  \BibitemOpen
  \bibfield  {author} {\bibinfo {author} {\bibfnamefont {B.}~\bibnamefont
  {D{\"u}nweg}}\ and\ \bibinfo {author} {\bibfnamefont {K.}~\bibnamefont
  {Kremer}},\ }\href@noop {} {\bibfield  {journal} {\bibinfo  {journal} {The
  Journal of chemical physics}\ }\textbf {\bibinfo {volume} {99}},\ \bibinfo
  {pages} {6983} (\bibinfo {year} {1993})}\BibitemShut {NoStop}%
\bibitem [{\citenamefont {Praprotnik}, \citenamefont {Site},\ and\
  \citenamefont {Kremer}(2008)}]{praprotnik2008multiscale}%
  \BibitemOpen
  \bibfield  {author} {\bibinfo {author} {\bibfnamefont {M.}~\bibnamefont
  {Praprotnik}}, \bibinfo {author} {\bibfnamefont {L.~D.}\ \bibnamefont
  {Site}}, \ and\ \bibinfo {author} {\bibfnamefont {K.}~\bibnamefont
  {Kremer}},\ }\href@noop {} {\bibfield  {journal} {\bibinfo  {journal} {Annu.
  Rev. Phys. Chem.}\ }\textbf {\bibinfo {volume} {59}},\ \bibinfo {pages} {545}
  (\bibinfo {year} {2008})}\BibitemShut {NoStop}%
\bibitem [{\citenamefont {Xue}\ \emph {et~al.}(2021)\citenamefont {Xue},
  \citenamefont {Biferale}, \citenamefont {Sbragaglia},\ and\ \citenamefont
  {Toschi}}]{xue2021lattice}%
  \BibitemOpen
  \bibfield  {author} {\bibinfo {author} {\bibfnamefont {X.}~\bibnamefont
  {Xue}}, \bibinfo {author} {\bibfnamefont {L.}~\bibnamefont {Biferale}},
  \bibinfo {author} {\bibfnamefont {M.}~\bibnamefont {Sbragaglia}}, \ and\
  \bibinfo {author} {\bibfnamefont {F.}~\bibnamefont {Toschi}},\ }\href@noop {}
  {\bibfield  {journal} {\bibinfo  {journal} {The European Physical Journal E}\
  }\textbf {\bibinfo {volume} {44}},\ \bibinfo {pages} {1} (\bibinfo {year}
  {2021})}\BibitemShut {NoStop}%
\bibitem [{\citenamefont {Parsa}\ and\ \citenamefont
  {Wagner}(2020)}]{parsa2020large}%
  \BibitemOpen
  \bibfield  {author} {\bibinfo {author} {\bibfnamefont {M.~R.}\ \bibnamefont
  {Parsa}}\ and\ \bibinfo {author} {\bibfnamefont {A.~J.}\ \bibnamefont
  {Wagner}},\ }\href@noop {} {\bibfield  {journal} {\bibinfo  {journal}
  {Physical Review Letters}\ }\textbf {\bibinfo {volume} {124}},\ \bibinfo
  {pages} {234501} (\bibinfo {year} {2020})}\BibitemShut {NoStop}%
\bibitem [{\citenamefont {Belardinelli}\ \emph {et~al.}(2015)\citenamefont
  {Belardinelli}, \citenamefont {Sbragaglia}, \citenamefont {Biferale},
  \citenamefont {Gross},\ and\ \citenamefont
  {Varnik}}]{belardinelli2015fluctuating}%
  \BibitemOpen
  \bibfield  {author} {\bibinfo {author} {\bibfnamefont {D.}~\bibnamefont
  {Belardinelli}}, \bibinfo {author} {\bibfnamefont {M.}~\bibnamefont
  {Sbragaglia}}, \bibinfo {author} {\bibfnamefont {L.}~\bibnamefont
  {Biferale}}, \bibinfo {author} {\bibfnamefont {M.}~\bibnamefont {Gross}}, \
  and\ \bibinfo {author} {\bibfnamefont {F.}~\bibnamefont {Varnik}},\
  }\href@noop {} {\bibfield  {journal} {\bibinfo  {journal} {Physical Review
  E}\ }\textbf {\bibinfo {volume} {91}},\ \bibinfo {pages} {023313} (\bibinfo
  {year} {2015})}\BibitemShut {NoStop}%
\bibitem [{\citenamefont {Gross}\ \emph {et~al.}(2011)\citenamefont {Gross},
  \citenamefont {Adhikari}, \citenamefont {Cates},\ and\ \citenamefont
  {Varnik}}]{gross2011modelling}%
  \BibitemOpen
  \bibfield  {author} {\bibinfo {author} {\bibfnamefont {M.}~\bibnamefont
  {Gross}}, \bibinfo {author} {\bibfnamefont {R.}~\bibnamefont {Adhikari}},
  \bibinfo {author} {\bibfnamefont {M.}~\bibnamefont {Cates}}, \ and\ \bibinfo
  {author} {\bibfnamefont {F.}~\bibnamefont {Varnik}},\ }\href@noop {}
  {\bibfield  {journal} {\bibinfo  {journal} {Philosophical Transactions of the
  Royal Society A: Mathematical, Physical and Engineering Sciences}\ }\textbf
  {\bibinfo {volume} {369}},\ \bibinfo {pages} {2274} (\bibinfo {year}
  {2011})}\BibitemShut {NoStop}%
\bibitem [{\citenamefont {Ladd}(1994)}]{ladd1994numerical}%
  \BibitemOpen
  \bibfield  {author} {\bibinfo {author} {\bibfnamefont {A.~J.}\ \bibnamefont
  {Ladd}},\ }\href@noop {} {\bibfield  {journal} {\bibinfo  {journal} {Journal
  of fluid mechanics}\ }\textbf {\bibinfo {volume} {271}},\ \bibinfo {pages}
  {285} (\bibinfo {year} {1994})}\BibitemShut {NoStop}%
\bibitem [{\citenamefont {Ladd}(1993)}]{ladd1993short}%
  \BibitemOpen
  \bibfield  {author} {\bibinfo {author} {\bibfnamefont {A.~J.}\ \bibnamefont
  {Ladd}},\ }\href@noop {} {\bibfield  {journal} {\bibinfo  {journal} {Physical
  Review Letters}\ }\textbf {\bibinfo {volume} {70}},\ \bibinfo {pages} {1339}
  (\bibinfo {year} {1993})}\BibitemShut {NoStop}%
\bibitem [{\citenamefont {D{\"u}nweg}\ and\ \citenamefont
  {Ladd}(2009)}]{dunweg2009lattice}%
  \BibitemOpen
  \bibfield  {author} {\bibinfo {author} {\bibfnamefont {B.}~\bibnamefont
  {D{\"u}nweg}}\ and\ \bibinfo {author} {\bibfnamefont {A.~J.}\ \bibnamefont
  {Ladd}},\ }in\ \href@noop {} {\emph {\bibinfo {booktitle} {Advanced computer
  simulation approaches for soft matter sciences III}}}\ (\bibinfo  {publisher}
  {Springer},\ \bibinfo {year} {2009})\ pp.\ \bibinfo {pages}
  {89--166}\BibitemShut {NoStop}%
\bibitem [{\citenamefont {D{\"u}nweg}, \citenamefont {Schiller},\ and\
  \citenamefont {Ladd}(2007)}]{dunweg2007statistical}%
  \BibitemOpen
  \bibfield  {author} {\bibinfo {author} {\bibfnamefont {B.}~\bibnamefont
  {D{\"u}nweg}}, \bibinfo {author} {\bibfnamefont {U.~D.}\ \bibnamefont
  {Schiller}}, \ and\ \bibinfo {author} {\bibfnamefont {A.~J.}\ \bibnamefont
  {Ladd}},\ }\href@noop {} {\bibfield  {journal} {\bibinfo  {journal} {Physical
  Review E—Statistical, Nonlinear, and Soft Matter Physics}\ }\textbf
  {\bibinfo {volume} {76}},\ \bibinfo {pages} {036704} (\bibinfo {year}
  {2007})}\BibitemShut {NoStop}%
\bibitem [{\citenamefont {Lauricella}\ \emph
  {et~al.}(2025{\natexlab{a}})\citenamefont {Lauricella}, \citenamefont
  {Montessori}, \citenamefont {Tiribocchi},\ and\ \citenamefont
  {Succi}}]{lauricella2025regularized}%
  \BibitemOpen
  \bibfield  {author} {\bibinfo {author} {\bibfnamefont {M.}~\bibnamefont
  {Lauricella}}, \bibinfo {author} {\bibfnamefont {A.}~\bibnamefont
  {Montessori}}, \bibinfo {author} {\bibfnamefont {A.}~\bibnamefont
  {Tiribocchi}}, \ and\ \bibinfo {author} {\bibfnamefont {S.}~\bibnamefont
  {Succi}},\ }\href@noop {} {\bibfield  {journal} {\bibinfo  {journal} {The
  Journal of Chemical Physics}\ }\textbf {\bibinfo {volume} {163}} (\bibinfo
  {year} {2025}{\natexlab{a}})}\BibitemShut {NoStop}%
\bibitem [{\citenamefont {Jacob}, \citenamefont {Malaspinas},\ and\
  \citenamefont {Sagaut}(2018)}]{jacob2018new}%
  \BibitemOpen
  \bibfield  {author} {\bibinfo {author} {\bibfnamefont {J.}~\bibnamefont
  {Jacob}}, \bibinfo {author} {\bibfnamefont {O.}~\bibnamefont {Malaspinas}}, \
  and\ \bibinfo {author} {\bibfnamefont {P.}~\bibnamefont {Sagaut}},\
  }\href@noop {} {\bibfield  {journal} {\bibinfo  {journal} {Journal of
  Turbulence}\ }\textbf {\bibinfo {volume} {19}},\ \bibinfo {pages} {1051}
  (\bibinfo {year} {2018})}\BibitemShut {NoStop}%
\bibitem [{\citenamefont {Coreixas}\ \emph {et~al.}(2017)\citenamefont
  {Coreixas}, \citenamefont {Wissocq}, \citenamefont {Puigt}, \citenamefont
  {Boussuge},\ and\ \citenamefont {Sagaut}}]{coreixas2017recursive}%
  \BibitemOpen
  \bibfield  {author} {\bibinfo {author} {\bibfnamefont {C.}~\bibnamefont
  {Coreixas}}, \bibinfo {author} {\bibfnamefont {G.}~\bibnamefont {Wissocq}},
  \bibinfo {author} {\bibfnamefont {G.}~\bibnamefont {Puigt}}, \bibinfo
  {author} {\bibfnamefont {J.-F.}\ \bibnamefont {Boussuge}}, \ and\ \bibinfo
  {author} {\bibfnamefont {P.}~\bibnamefont {Sagaut}},\ }\href@noop {}
  {\bibfield  {journal} {\bibinfo  {journal} {Physical Review E}\ }\textbf
  {\bibinfo {volume} {96}},\ \bibinfo {pages} {033306} (\bibinfo {year}
  {2017})}\BibitemShut {NoStop}%
\bibitem [{\citenamefont {Mattila}, \citenamefont {Philippi},\ and\
  \citenamefont {Hegele}(2017)}]{mattila2017high}%
  \BibitemOpen
  \bibfield  {author} {\bibinfo {author} {\bibfnamefont {K.~K.}\ \bibnamefont
  {Mattila}}, \bibinfo {author} {\bibfnamefont {P.~C.}\ \bibnamefont
  {Philippi}}, \ and\ \bibinfo {author} {\bibfnamefont {L.~A.}\ \bibnamefont
  {Hegele}},\ }\href@noop {} {\bibfield  {journal} {\bibinfo  {journal}
  {Physics of Fluids}\ }\textbf {\bibinfo {volume} {29}} (\bibinfo {year}
  {2017})}\BibitemShut {NoStop}%
\bibitem [{\citenamefont {Latt}\ and\ \citenamefont
  {Chopard}(2006)}]{latt2006lattice}%
  \BibitemOpen
  \bibfield  {author} {\bibinfo {author} {\bibfnamefont {J.}~\bibnamefont
  {Latt}}\ and\ \bibinfo {author} {\bibfnamefont {B.}~\bibnamefont {Chopard}},\
  }\href@noop {} {\bibfield  {journal} {\bibinfo  {journal} {Mathematics and
  Computers in Simulation}\ }\textbf {\bibinfo {volume} {72}},\ \bibinfo
  {pages} {165} (\bibinfo {year} {2006})}\BibitemShut {NoStop}%
\bibitem [{\citenamefont {Schiller}(2008)}]{schiller2008thermal}%
  \BibitemOpen
  \bibfield  {author} {\bibinfo {author} {\bibfnamefont {U.~D.}\ \bibnamefont
  {Schiller}},\ }\emph {\bibinfo {title} {Thermal fluctuations and boundary
  conditions in the lattice Boltzmann method}},\ \href@noop {} {Ph.D. thesis},\
  \bibinfo  {school} {Johannes Gutenberg Universit{\"a}t Mainz} (\bibinfo
  {year} {2008})\BibitemShut {NoStop}%
\bibitem [{\citenamefont {Malaspinas}(2015)}]{malaspinas2015increasing}%
  \BibitemOpen
  \bibfield  {author} {\bibinfo {author} {\bibfnamefont {O.}~\bibnamefont
  {Malaspinas}},\ }\href@noop {} {\bibfield  {journal} {\bibinfo  {journal}
  {arXiv preprint arXiv:1505.06900}\ } (\bibinfo {year} {2015})}\BibitemShut
  {NoStop}%
\bibitem [{\citenamefont {Montessori}\ \emph {et~al.}(2023)\citenamefont
  {Montessori}, \citenamefont {Lauricella}, \citenamefont {Tiribocchi},
  \citenamefont {Durve}, \citenamefont {La~Rocca}, \citenamefont {Amati},
  \citenamefont {Bonaccorso},\ and\ \citenamefont
  {Succi}}]{montessori2023thread}%
  \BibitemOpen
  \bibfield  {author} {\bibinfo {author} {\bibfnamefont {A.}~\bibnamefont
  {Montessori}}, \bibinfo {author} {\bibfnamefont {M.}~\bibnamefont
  {Lauricella}}, \bibinfo {author} {\bibfnamefont {A.}~\bibnamefont
  {Tiribocchi}}, \bibinfo {author} {\bibfnamefont {M.}~\bibnamefont {Durve}},
  \bibinfo {author} {\bibfnamefont {M.}~\bibnamefont {La~Rocca}}, \bibinfo
  {author} {\bibfnamefont {G.}~\bibnamefont {Amati}}, \bibinfo {author}
  {\bibfnamefont {F.}~\bibnamefont {Bonaccorso}}, \ and\ \bibinfo {author}
  {\bibfnamefont {S.}~\bibnamefont {Succi}},\ }\href@noop {} {\bibfield
  {journal} {\bibinfo  {journal} {Journal of Computational Science}\ }\textbf
  {\bibinfo {volume} {74}},\ \bibinfo {pages} {102165} (\bibinfo {year}
  {2023})}\BibitemShut {NoStop}%
\bibitem [{\citenamefont {Montessori}\ \emph {et~al.}(2024)\citenamefont
  {Montessori}, \citenamefont {La~Rocca}, \citenamefont {Amati}, \citenamefont
  {Lauricella}, \citenamefont {Tiribocchi},\ and\ \citenamefont
  {Succi}}]{montessori2024high}%
  \BibitemOpen
  \bibfield  {author} {\bibinfo {author} {\bibfnamefont {A.}~\bibnamefont
  {Montessori}}, \bibinfo {author} {\bibfnamefont {M.}~\bibnamefont
  {La~Rocca}}, \bibinfo {author} {\bibfnamefont {G.}~\bibnamefont {Amati}},
  \bibinfo {author} {\bibfnamefont {M.}~\bibnamefont {Lauricella}}, \bibinfo
  {author} {\bibfnamefont {A.}~\bibnamefont {Tiribocchi}}, \ and\ \bibinfo
  {author} {\bibfnamefont {S.}~\bibnamefont {Succi}},\ }\href@noop {}
  {\bibfield  {journal} {\bibinfo  {journal} {Physics of Fluids}\ }\textbf
  {\bibinfo {volume} {36}} (\bibinfo {year} {2024})}\BibitemShut {NoStop}%
\bibitem [{\citenamefont {Lauricella}\ \emph
  {et~al.}(2025{\natexlab{b}})\citenamefont {Lauricella}, \citenamefont
  {Mukherjee}, \citenamefont {Brandt}, \citenamefont {Succi}, \citenamefont
  {Izbassarov},\ and\ \citenamefont {Montessori}}]{lauricella2025acclb}%
  \BibitemOpen
  \bibfield  {author} {\bibinfo {author} {\bibfnamefont {M.}~\bibnamefont
  {Lauricella}}, \bibinfo {author} {\bibfnamefont {A.}~\bibnamefont
  {Mukherjee}}, \bibinfo {author} {\bibfnamefont {L.}~\bibnamefont {Brandt}},
  \bibinfo {author} {\bibfnamefont {S.}~\bibnamefont {Succi}}, \bibinfo
  {author} {\bibfnamefont {D.}~\bibnamefont {Izbassarov}}, \ and\ \bibinfo
  {author} {\bibfnamefont {A.}~\bibnamefont {Montessori}},\ }\href@noop {}
  {\bibfield  {journal} {\bibinfo  {journal} {arXiv preprint arXiv:2505.01126}\
  } (\bibinfo {year} {2025}{\natexlab{b}})}\BibitemShut {NoStop}%
\bibitem [{\citenamefont {Landau}\ and\ \citenamefont
  {Lifshitz}(1987)}]{landau1987fluid}%
  \BibitemOpen
  \bibfield  {author} {\bibinfo {author} {\bibfnamefont {L.~D.}\ \bibnamefont
  {Landau}}\ and\ \bibinfo {author} {\bibfnamefont {E.~M.}\ \bibnamefont
  {Lifshitz}},\ }\href@noop {} {\emph {\bibinfo {title} {Fluid Mechanics:
  Volume 6}}},\ Vol.~\bibinfo {volume} {6}\ (\bibinfo  {publisher} {Elsevier},\
  \bibinfo {year} {1987})\BibitemShut {NoStop}%
\bibitem [{\citenamefont {Ollila}\ \emph {et~al.}(2011)\citenamefont {Ollila},
  \citenamefont {Denniston}, \citenamefont {Karttunen},\ and\ \citenamefont
  {Ala-Nissila}}]{ollila2011fluctuating}%
  \BibitemOpen
  \bibfield  {author} {\bibinfo {author} {\bibfnamefont {S.~T.}\ \bibnamefont
  {Ollila}}, \bibinfo {author} {\bibfnamefont {C.}~\bibnamefont {Denniston}},
  \bibinfo {author} {\bibfnamefont {M.}~\bibnamefont {Karttunen}}, \ and\
  \bibinfo {author} {\bibfnamefont {T.}~\bibnamefont {Ala-Nissila}},\
  }\href@noop {} {\bibfield  {journal} {\bibinfo  {journal} {The Journal of
  chemical physics}\ }\textbf {\bibinfo {volume} {134}} (\bibinfo {year}
  {2011})}\BibitemShut {NoStop}%
\bibitem [{\citenamefont {Bernaschi}\ \emph {et~al.}(2009)\citenamefont
  {Bernaschi}, \citenamefont {Melchionna}, \citenamefont {Succi}, \citenamefont
  {Fyta}, \citenamefont {Kaxiras},\ and\ \citenamefont
  {Sircar}}]{bernaschi2009muphy}%
  \BibitemOpen
  \bibfield  {author} {\bibinfo {author} {\bibfnamefont {M.}~\bibnamefont
  {Bernaschi}}, \bibinfo {author} {\bibfnamefont {S.}~\bibnamefont
  {Melchionna}}, \bibinfo {author} {\bibfnamefont {S.}~\bibnamefont {Succi}},
  \bibinfo {author} {\bibfnamefont {M.}~\bibnamefont {Fyta}}, \bibinfo {author}
  {\bibfnamefont {E.}~\bibnamefont {Kaxiras}}, \ and\ \bibinfo {author}
  {\bibfnamefont {J.~K.}\ \bibnamefont {Sircar}},\ }\href@noop {} {\bibfield
  {journal} {\bibinfo  {journal} {Computer Physics Communications}\ }\textbf
  {\bibinfo {volume} {180}},\ \bibinfo {pages} {1495} (\bibinfo {year}
  {2009})}\BibitemShut {NoStop}%
\bibitem [{\citenamefont {d'Humi{\`e}res}(2002)}]{d2002multiple}%
  \BibitemOpen
  \bibfield  {author} {\bibinfo {author} {\bibfnamefont {D.}~\bibnamefont
  {d'Humi{\`e}res}},\ }\href@noop {} {\bibfield  {journal} {\bibinfo  {journal}
  {Philosophical Transactions of the Royal Society of London. Series A:
  Mathematical, Physical and Engineering Sciences}\ }\textbf {\bibinfo {volume}
  {360}},\ \bibinfo {pages} {437} (\bibinfo {year} {2002})}\BibitemShut
  {NoStop}%
\bibitem [{\citenamefont {Lallemand}\ and\ \citenamefont
  {Luo}(2000)}]{lallemand2000theory}%
  \BibitemOpen
  \bibfield  {author} {\bibinfo {author} {\bibfnamefont {P.}~\bibnamefont
  {Lallemand}}\ and\ \bibinfo {author} {\bibfnamefont {L.-S.}\ \bibnamefont
  {Luo}},\ }\href@noop {} {\bibfield  {journal} {\bibinfo  {journal} {Physical
  review E}\ }\textbf {\bibinfo {volume} {61}},\ \bibinfo {pages} {6546}
  (\bibinfo {year} {2000})}\BibitemShut {NoStop}%
\end{thebibliography}

%merlin.mbs aipnum4-1.bst 2010-07-25 4.21a (PWD, AO, DPC) hacked
%Control: key (0)
%Control: author (8) initials jnrlst
%Control: editor formatted (1) identically to author
%Control: production of article title (-1) disabled
%Control: page (0) single
%Control: year (1) truncated
%Control: production of eprint (0) enabled
%

\end{document}